\documentclass[draftcls,onecolumn,12pt]{IEEEtran}
\usepackage{bm}
\usepackage{epsfig}
\usepackage{stfloats}
\usepackage{graphicx}
\usepackage{subfigure}
\usepackage{amssymb}

\usepackage{epstopdf}
\usepackage[sort, compress]{cite}
\usepackage{epsfig,graphics,subfigure}
\usepackage[cmex10]{amsmath}

\usepackage{array}
\usepackage{tabulary}
\usepackage{booktabs}
\usepackage{amsfonts}
\usepackage{algorithmic}
\usepackage{algorithm}

\begin{document}

\bibliographystyle{IEEEtran}

\title{Cooperative Jamming for Secure Transmission With Both Active and Passive Eavesdroppers}
\author{\IEEEauthorblockN{Jiangbo Si, {\emph{Member, IEEE}}, Zihao Cheng, Zan Li, {\emph{Senior Member, IEEE}},
Julian Cheng, {\emph{Senior Member, IEEE}}, Hui-Ming Wang, {\emph{Senior Member, IEEE}}, and Naofal Al-Dhahir, \emph{Fellow, IEEE}}
\thanks{}
\thanks{Jiangbo Si, Zihao Cheng, and Zan Li are with the Integrated Service Networks Lab
of Xidian University, Xi'an, 710071, China. Jiangbo Si is also a visiting scholar at the University of Texas at Dallas. (e-mail: jbsi@xidian.edu.cn).}
\thanks{Julian Cheng is with the School of Engineering, The University
of British Columbia, Kelowna, BC V1V 1V7, Canada (e-mail: julian.cheng@ubc.ca).}
\thanks{Hui-Ming Wang is with the School of Electronic and Information Engineering,
Xi¡¯an Jiaotong University, Xi¡¯an 710049, China (e-mail: xjbswhm@gmail.com).}
\thanks{N. Al-Dhahir is with the Department of Electrical and Computer Engineering, The University of Texas at Dallas, Richardson,
TX 75080 USA.(e-mail: aldhahir@utdallas.edu).}
}
 \maketitle
\begin{abstract}
  Secrecy transmission is investigated for a cooperative jamming scheme, where a multi-antenna jammer generates artificial noise (AN) to confuse eavesdroppers. Two kinds of eavesdroppers are considered: passive eavesdroppers who only overhear the legitimate information, and active eavesdroppers who not only overhear the legitimate information but also jam the legitimate signal. Existing works only treat the passive and active eavesdroppers separately. Different from the existing works, we investigate the achievable secrecy rate in presence of both active and passive eavesdroppers. For the considered system model, we assume that the instantaneous channel state information (CSI) of the active eavesdroppers is available at the jammer, while only partial CSI of the passive eavesdroppers is available at the jammer. A new zero-forcing beamforming scheme is proposed in the presence of both active and passive eavesdroppers. For both the perfect and imperfect CSI cases, the total transmission power allocation between the information and AN signals is optimized to maximize the achievable secrecy rate. Numerical results show that imperfect CSI between the jammer and the legitimate receiver will do more harm to the achievable secrecy rate than imperfect CSI between the jammer and the active eavesdropper.
\end{abstract}
\IEEEpeerreviewmaketitle
\section{Introduction}
Physical layer security (PLS) techniques have attracted tremendous research attention in the past decade. Different from the conventional encryption methods, PLS can achieve positive secrecy rate without a predetermined secrecy key \cite{Bloch}. Many approaches have been proposed to enhance the secrecy performance of PLS. For example,  multiple antennas and relay techniques are often deployed to enhance the reception performance at the legitimate receiver. On the other hand, artificial noise (AN) and cooperative jamming can be deployed to degrade the reception performance at the eavesdropper. When there is a high secrecy rate requirement or when the eavesdropper channel quality is better than the legitimate channel quality, AN and cooperative jamming become attractive schemes to secure transmission \cite{HWang1,NZhao,DWang}.

A plethora of works have investigated beamforming with AN under different secrecy goals and different transmission scenarios. For instance, when transmit beamforming is deployed at the multi-antenna transmitter, the achievable secrecy rate \cite{XZhou1,AAl-Naharicognitive}, the secrecy outage probability (SOP) \cite{TZheng}, and the secrecy throughput\cite{XZhou2,CWang2} can be improved by optimizing the power allocation between the information signal and AN. Under both secrecy and transmission outage constraints, an on-off AN scheme was investigated to maximize the effective secrecy throughput in \cite{NYang3,NYang4}, where fully adaptive AN schemes, i.e, adaptive transmission rate and adaptive secrecy rate, were proposed. By contrast, a partial adaptive AN scheme was proposed in \cite{SYan} for a fixed secrecy rate and a varying transmission rate. The results in \cite{SYan} indicate that the partial adaptive scheme outperforms the on-off scheme, and can achieve almost the same secrecy performance as the fully adaptive scheme. In addition, when a normal user and a secure user \footnote{A normal user transmits public messages without a secrecy requirement, and a secure user transmits privacy messages with a secrecy constraint.} coexist in a cellular network \cite{WWang1}, under the constraint of average throughput for the normal user, the effective secrecy throughputs for the non-adaptive and fully adaptive AN schemes were maximized by the optimal power allocation among the secure user, AN, and the normal user. In addition, due to feedback overhead \cite{JHu}, time-varying channels \cite{HYu} and imperfect channel state information (CSI) estimation \cite{XZhou1,TZheng}, AN towards the eavesdroppers can be leaked to the legitimate receiver. Moreover, for the case of imperfect CSI, optimal power allocation between the legitimate information signal and AN was also studied in \cite{XZhou1,TZheng} for high reliability and security transmission.

When multiple antennas are not available at the transmitter and the AN technique cannot be used, a friendly jammer can be employed to degrade the eavesdropper channel. Without the eavesdroppers' instantaneous CSI, a joint jamming and beamforming design was considered in \cite{LHu1,VNguyencognitive}, where the jammer equipped with beamforming injects AN into the null space of the legitimate receiver. Similarly, multiple single antenna jammers can cooperatively form the transmit beamforming and inject AN towards the eavesdroppers \cite{CWang1,HGuo}. These works assumed that transmit beamforming was applied at a multi-antenna jammer or multiple single-antenna jammers. Different from these works, a new jamming scheme was proposed without transmit beamforming in \cite{KCumanan}, while multiple jammers cooperatively transmit the noise towards the legitimate receiver and the eavesdroppers. Moreover, to reduce the system complexity, one `best' jammer that has the minimum noise power at the legitimate receiver was selected to forward the AN under both perfect and imperfect CSI  assumptions \cite{LWang}.


The aforementioned works assumed that the eavesdroppers' instantaneous CSIs were unavailable at the legitimate transmitter and the jammer. These eavesdroppers can be considered as the passive eavesdroppers, who overhear legitimate messages silently. However, besides the passive eavesdropper, recently several works have considered the active eavesdropper who can transmit malicious jamming
signals and receive legitimate signals simultaneously \cite{CLiu,YWu,LLi,LKong,ZLiu}. Similarly, when multiple antennas were deployed at the active eavesdropper and legitimate receiver, the power minimization and secrecy rate maximization problems were solved in \cite{ZChu,HWu}. In addition, to circumvent the imperfect CSI issues, robust techniques were introduced in \cite{JHuang,HMa,QLi}, where cooperative jamming schemes were designed based on the worst-case secrecy rate.

In practice, the eavesdroppers can cooperate to overhear the legitimate information. For example, in a wireless network, active eavesdroppers may intentionally expose themselves to attract the legitimate user's attention by injecting AN. Simultaneously, passive eavesdroppers overhear the legitimate information silently.  Considering the active eavesdroppers or the passive eavesdroppers alone cannot achieve the maximum secrecy rate \cite{HZhang,GGeraci} and the secure performance cannot be guaranteed. Hence, for the legitimate user, it is necessary to design the secrecy transmission scheme considering both the active eavesdropper and passive eavesdropper. To the best of the authors' knowledge, there is no prior work focusing on secrecy rate maximization when both active and passive eavesdroppers coexist in a wireless network. Hence, we propose a cooperative jamming scheme to achieve high reliability and security transmission for a wireless network having coexisting active and passive eavesdroppers. Both the instantaneous CSI of the active eavesdropper link and the partial CSI of the passive eavesdroppers links are exploited to improve the secrecy performance. The main contributions of this paper are summarized as follows:

\begin{itemize}
\item[1.] For the scenario when the active and passive eavesdroppers coexist in a wireless network, we propose a cooperative two-fold zero-forcing jamming scheme for secrecy transmission. In the proposed scheme, a multi-antenna jammer treats the active and passive eavesdroppers differently. Specifically, maximal ratio transmission (MRT) is used as the beamforming vector towards the active eavesdropper, and the AN for the active eavesdropper lies in the null space of the legitimate receiver. By contrast, a random beamforming vector is used for the passive eavesdropper, and the AN for the passive eavesdroppers lies in the null space of both the legitimate receiver and the active eavesdroppers. Since both the instantaneous CSI of the active eavesdropper link and partial CSI of the passive eavesdropper links are exploited, the proposed jamming scheme can significantly improve the secrecy rate over the conventional beamforming with AN schemes that only take the passive eavesdroppers into account, and inject AN equally towards all directions.

\item[2.] Under both the transmission reliability and security constraints, the total transmission power is allocated among the legitimate information signal, AN for the active eavesdropper, and AN for the passive eavesdroppers to maximize the secrecy rate. The legitimate information power is first derived after we reveal a monotonicity relationship in transmission outage probability and SOP. Then, we prove the monotonicity of SOP with the AN power allocated to the active eavesdropper, concavity of SOP with the AN power allocated to the passive eavesdroppers, and the monotonicity of SOP with the secrecy rate. Based on these properties, the secrecy rate is maximized by optimally allocating the remaining power to AN for the active eavesdropper and AN for the passive eavesdroppers, respectively.

\item[3.] In practice, the instantaneous CSI can be imperfect due to feedback delay or estimation error. To investigate the impact of imperfect CSI on secrecy performance, we derive exact expressions for the transmission outage probability and SOP under imperfect CSI. Using these expressions, we study the relationship between the power allocation ratio and SOP, and obtain the maximum secrecy rate as well as optimal design parameters. We reveal that imperfect CSI between the jammer and Bob always reduces the achieved maximum secrecy rate. By contrast, the imperfect CSI of the jammer to the active eavesdropper link does not affect the achieved maximum secrecy rate when the channel quality between the jammer and the active eavesdropper is high.
\end{itemize}

The remainder of this paper is organized as follows.
Section II describes the system model and presents a cooperative jamming scheme for secrecy transmission. Section III formulates an optimization problem to maximize the secrecy rate, and proposes a numerical method to solve the optimization problem. In Section IV, optimizing the transmit power allocation between the information signal and AN is investigated with imperfect CSI. In Section V, the cooperative jamming scheme is proposed when there are multiple active eavesdroppers in the network.  Numerical results are presented and discussed in Section VI. Finally, we conclude the paper in Section VII.

\emph{Notations}- $\left(  \cdot  \right)^T$, $\left(  \cdot  \right)^*$ and $\left(  \cdot  \right)^H$ denote the transpose, conjugate, and conjugate transpose, respectively; $\left\| {\left.  \cdot  \right\|} \right.$ denotes the Frobenius norm. $f_\upsilon\left(  \cdot  \right)$ is the probability density function (PDF) of  random variable (RV) $\upsilon$; $F_\upsilon\left(  \cdot  \right)$ is the cumulative distribution
function (CDF) of RV $\upsilon$; ${\phi_\upsilon}\left( s \right)$ is the moment generation function (MGF) of $\upsilon$; $\mathbb{E}{(\cdot)}$ is the expected value of a RV; $Gamma(\mu, \sigma^2)$ denotes the gamma distribution with
shape parameter $\mu$ and scale parameter $\sigma^2$; $\exp(\sigma^2)$ denotes the exponential distribution with mean $\sigma^2$. $
\chi ^2 \left( {{\nu}} \right)$ denotes the central chi-square distribution with $\nu$ degrees of freedom;
${{{\mathbb{C}}}^{m \times n}}$ denotes the set of ${m \times n}$ complex matrices; ${\bf{x}} \sim \mathcal{CN}\left( {{\bf{\Lambda }},{\bf{\Delta }}} \right)$  denotes the circular symmetric
complex Gaussian vector with mean vector $\bf{\Lambda }$ and covariance matrix ${\bf{\Delta }}$.
\section{System Model}
\begin{figure}[ht]
\centering
\includegraphics[width=3.5in]{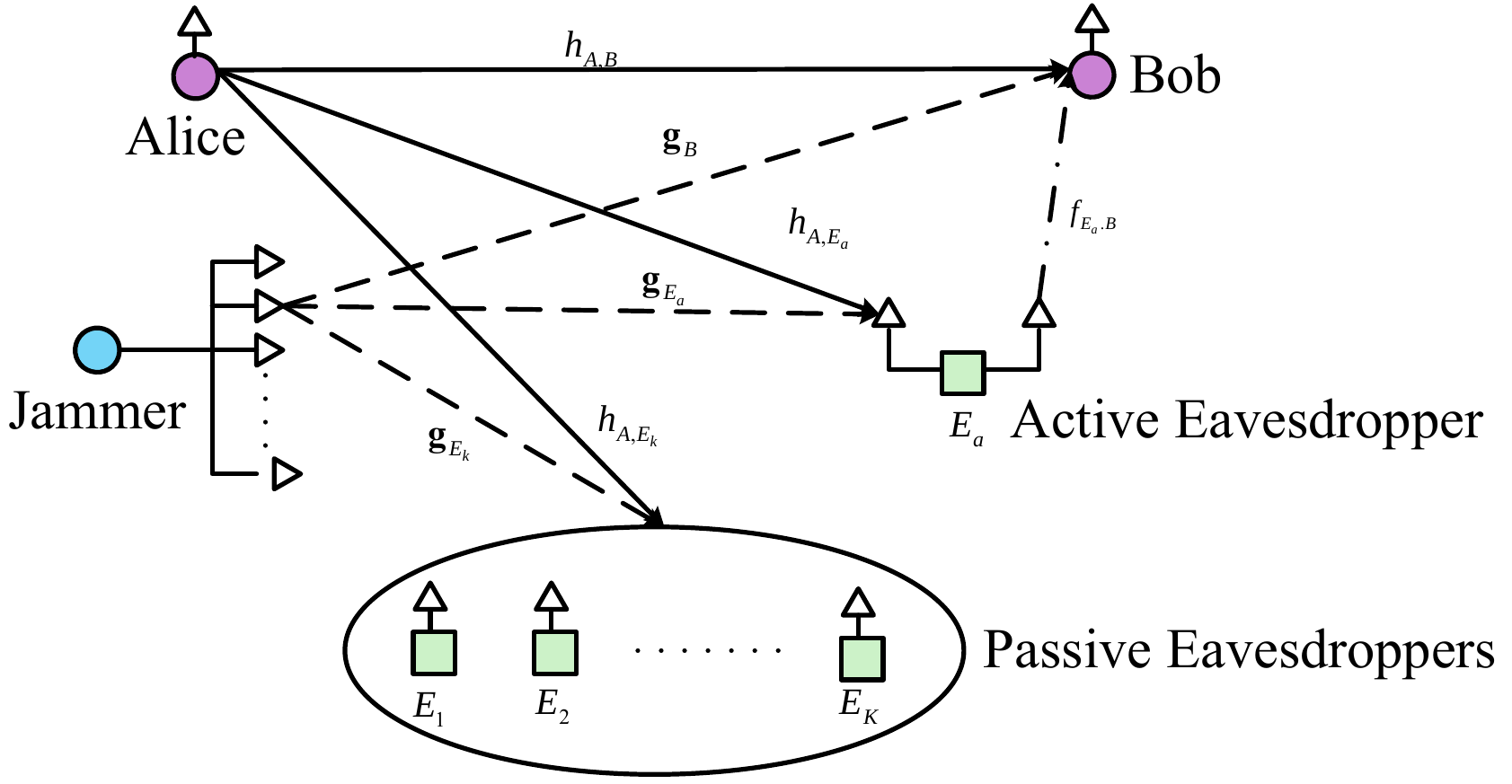}
\caption{System model of secure communication systems with both the active and passive eavesdroppers} \label{fig.1}
\end{figure}
We consider the wireless network shown in Fig. 1, where the transmitter (Alice) aims to establish a secure communication link with the legitimate receiver (Bob) in the presence of an active eavesdropper $E_{a}$ and $K$ passive eavesdroppers $E_{k}$ $(1\leq k \leq K)$, who attempt to overhear the legitimate information transmitted from Alice. The active eavesdropper equipped with two antennas operates in a full-duplex mode, where it transmits the AN towards the legitimate receiver and overhears the legitimate messages simultaneously. By contrast, the passive eavesdroppers overhear the legitimate messages silently. Moreover, to improve the secrecy performance, Alice deploys a friendly jammer $J$ equipped with $N$ antennas to confuse both the active and passive eavesdroppers. Due to the space and cost limitations, Alice, Bob and the passive eavesdroppers are each equipped with a single antenna, which is a reasonable assumption in device-to-device (D2D) networks.

All channels are assumed to experience Rayleigh flat fading. The instantaneous CSI of the Alice $\rightarrow$ Bob, Alice $\rightarrow$ $E_a$, Alice $\rightarrow$ $E_k$, and $E_a$ $\rightarrow$ Bob links are, respectively, denoted by the channel coefficients $h_{A,B}$, $h_{A,E_a}$, $h_{A,E_k}$, and $f_{E_a,B}$, which are complex Gaussian RVs having mean zero and variances $\sigma^2_{A,B}$, $\sigma^2_{A,E_a}$, $\sigma^2_{A,E_k}$, and $\sigma^2_{E_a,B}$, respectively. In addition, ${{\bf{g}}_{B}}\in \mathbb{C}^{N\times 1} \sim \mathcal {CN} \left( {{\bf{0}},{\sigma^2_{J,B}}{{\bf{I}}_{N}}} \right)$, ${\bf{g}}_{E_a}\in \mathbb{C}^{N\times 1} \sim \mathcal {CN} \left( {{\bf{0}},{\sigma^2_{J,E_a}}{{\bf{I}}_{N}}} \right)$, and ${\bf{g}}_{E_k}\in \mathbb{C}^{N\times 1} \sim \mathcal {CN} \left( {{\bf{0}},{\sigma^2_{J,E_k}}{{\bf{I}}_{N}}} \right)$ denote the instantaneous CSI of the jammer $\rightarrow$ Bob, the jammer $\rightarrow$ $E_a$, and the jammer $\rightarrow$ $E_k$ links, respectively. We assume that the instantaneous CSI of ${\bf{g}}_{E_a}$ ,  and the statistical CSI of ${\bf{g}}_{E_k}$ are available at Alice and the jammer, which is a reasonable assumption since the active eavesdropper injects AN towards Bob, while the passive eavesdroppers overhear the message silently.

\section{Optimal Power Allocation for Secrecy Rate Maximization}
There are two main schemes for secrecy rate maximization, namely, adaptive schemes and non-adaptive schemes \cite{WWang1,XZhou3}. Though an adaptive scheme can make full use of the wireless channel and improve the secrecy capacity, it increases the receiver complexity due to frequent change in the secrecy code rate. Moreover, since an adaptive scheme requires continuous change of the transmit power and thus has a higher complexity, we adopt a more practical and low-complexity non-adaptive scheme where the transmission rate $R_b$ and secrecy rate $R_s$ are fixed during the transmission. For the non-adaptive scheme, it is challenging to jointly optimize the beamforming coefficients and power allocation under a secrecy outage constraint. We will therefore first design a suboptimal beamforming scheme in the presence of both passive and active eavesdroppers. Then, based on the proposed beamforming scheme, we investigate the optimal power allocation between the legitimate signal, the AN for the active eavesdropper and the AN for the passive eavesdroppers.
\subsection{A Two-fold Zero-forcing Beamforming Scheme}
In the presence of both active and passive eavesdroppers, we propose a two-fold zero-forcing beamforming applied at the multi-antenna jammer $J$. More specifically, the AN injected towards the active eavesdropper will not interfere with Bob. In addition, the AN injected towards the passive eavesdropper will not interfere with Bob, as well as the active eavesdropper, because a separate AN signal has already been injected towards it based on instantaneous CSI between Jammer and the active eavesdropper. The proposed scheme makes a full use of the instantaneous CSI of the jammer $\rightarrow$ Bob and the jammer $\rightarrow$ $E_a$ links. Specifically, the jammer adopts MRT to maximize the AN power at the active eavesdropper. Thus, the beamforming vector for the active eavesdropper ${{{{{\mbox{\small\bf{W}}}}}_{{E_a}}}}\in \mathbb{C}^{N\times 1}$  should lie in the null-space of ${\bf{g}}_B$, and it is given by
\begin{align}\label{27}\
{{{{{\mbox{\small\bf{W}}}}}_{{E_a}}}} = \frac{{\left( {{{\bf{I}}_N} - \frac{{{{\bf{g}}_B}{{\bf{g}}^H}_B}}{{{{\left\| {{{\bf{g}}_B}} \right\|}^2}}}} \right){{\bf{g}}^*}_{{E_a}}}}{{\left\| {\left( {{{\bf{I}}_N} - \frac{{{{\bf{g}}_B}{{\bf{g}}^H}_B}}{{{{\left\| {{{\bf{g}}_B}} \right\|}^2}}}} \right){{\bf{g}}^*}_{{E_a}}} \right\|}}.
\end{align}
At the same time, since there are multiple passive eavesdroppers, the beamforming matrix for the passive eavesdroppers, ${{{\bf{W}}}_{{E_p}}}\in \mathbb{C}^{N\times(N-2)}$, must be designed to enhance secrecy performance. On the one hand, to avoid interfering with Bob, ${{{\bf{W}}}_{{E_p}}}$ should also lie in the null space of ${{\bf{g}}_B}$. On the other hand, since we have precisely injected AN to the active eavesdropper by MRT, beamforming for the passive eavesdroppers should remove the dimension corresponding to the active eavesdropper. Hence, ${{{\bf{W}}}_{{E_p}}}$ should lie in the null-space of $\left[{{\bf{g}}_B},{{{{\bf{\mbox{\small \bf{W}}}}}_{{E_a}}}}\right]$. Moreover, since the instantaneous CSI of the jammer $\rightarrow$ $E_k$ link is unavailable, the AN power for the passive eavesdroppers, $P_{J_P}$, is uniformly allocated to $N-2$ dimensions\cite{JXiong}. Thus the transmitted AN vector ${\bf{n}}_J\in \mathbb{C}^{N\times 1}$ is given by
\begin{align}\label{27}\
{{\bf{n}}_J} = {{{\sqrt{P_{ J_A}}{{\mbox{\small{\bf {W}}}}}_{{E_a}}}}}{{{n}}_{{J_A}}} + {{\sqrt{\frac{P_{J_P}}{N-2}}{\bf{W}}_{{E_p}}}}{{\bf{n}}_{{J_p}}}
\end{align}
where $P_{J_A}$ denotes the AN power for the active eavesdropper. In addition, ${n_{{J_A}}} \sim \mathcal {CN} \left( {0,1} \right)$ and ${{\bf{n}}_{J_p}}\in \mathbb{C}^{{\left( N - 2 \right)}\times 1} \sim \mathcal {CN} \left( {{\bf{0}},{{\bf{I}}_{N - 2}}} \right)$ denote the injected AN for the active eavesdropper and the passive eavesdroppers, respectively. Then, the received signals at the active eavesdropper $E_a$ and the passive eavesdropper $E_k$ are, respectively, given by
\begin{align}\label{signalEA}\
{{y}}_{E_a}  = \sqrt{P_A} {{h}}_{A,E_a} x + {{{\sqrt{P_{ J_A}}{{\bf{g}}^H_{E_a}}{{\mbox{\small{\bf {W}}}}}_{{E_a}}}}}{{{n}}_{{J_A}}}+ {{n}}_{E_a}
\end{align}
and
\begin{align}\label{signalEk}\
{{y}}_{E_k}  &= \sqrt{P_A} {{h}}_{A,E_k} x + {{{\sqrt{P_{ J_A}}{{\bf{g}}^H_{E_k}}{{\mbox{\small{\bf {W}}}}}_{{E_a}}}}}{{{n}}_{{J_A}}}\nonumber \\
&+ {{\sqrt{\frac{P_{J_P}}{N-2}}{{\bf{g}}^H_{E_k}}{\bf{W}}_{{E_p}}}}{{\bf{n}}_{{J_p}}}+ {{n}}_{E_k}, 1 \leq k \leq K
\end{align}
where $P_A$ is the transmit power at Alice. In addition, ${{n}}_{E_a}\sim \mathcal{CN}(0,1)$ and ${{n}}_{E_k}\sim \mathcal{CN}(0,1)$, respectively, denote the noises at the active eavesdropper $E_a$ and the passive eavesdropper $E_k$. Since all the eavesdroppers cooperatively overhear the legitimate information, the interference generated by $E_a$ is assumed to be perfectly cancelled at $E_k$ and $E_a$ by using advanced analog and digital interference cancellation methods \cite{JXu,JXu1};

Now, we derive the CDFs of signal-to-noise ratio (SNR) at the active eavesdropper, denoted as $\gamma_{E_a}$ and the passive eavesdroppers, denoted as $\gamma_{E_k} \left(1\leq k \leq K \right)$. Assume that the interference generated by the jammer dominates the noise power at the active eavesdropper, i.e.,  $\sigma^2_{J,E_a}\gg 1$.
 Thus, according to \eqref{signalEA},  ${\gamma _{{E_a}}}$ can be written as
 \begin{align}\label{gammaEA}\
{\gamma _{{E_a}}}& = \frac{{{P_A}{{\left| {{h_{A,{E_a}}}} \right|}^2}}}{{{P_{{J_A}}}{{\left| {{{\bf{g}}^H}_{{E_a}}{{{\mbox{\small{\bf{W}}}}}_{{E_a}}}}
\right|}^2}}}\nonumber \\
&= \frac{{{P_A}{\sigma ^2}_{A,{E_a}}{{{\lambda _1}} \mathord{\left/
 {\vphantom {{{\lambda _1}} 2}} \right.
 \kern-\nulldelimiterspace} 2}}}{{{P_{{J_A}}}{\sigma ^2}_{J,{E_a}}\left( {N - 1} \right){{{\lambda _2}} \mathord{\left/
 {\vphantom {{{\lambda _2}} {\left( {2(N - 1)} \right)}}} \right.
 \kern-\nulldelimiterspace} {\left( {2(N - 1)} \right)}}}}
\end{align}
where ${\left|{{h_{A,{E_a}}}} \right|^2} \sim Gamma\left( {1,{\sigma ^2}_{A,{E_a}}} \right)$; ${{\lambda _1} {\triangleq} {\frac{{{{\left| {{h_{A,{E_a}}}} \right|}^2}}}{{{\sigma ^2}_{A,{E_a}}/2}}}} \sim \chi^2(2)$; ${\lambda _2} \triangleq {\frac{{{{\left| {{g^H}_{{E_a}}{{{{\mbox{\small{\bf{W}}}}}_{{E_a}}}}} \right|}^2}}}{{{\sigma ^2}_{J,{E_a}}/2}}} \sim \chi^2(2(N-1))$. Thus, the ratio $\frac{{{{{\lambda _1}}
\mathord{\left/
 {\vphantom {{{\lambda _1}} 2}} \right.
 \kern-\nulldelimiterspace} 2}}}{{{{{\lambda _2}} \mathord{\left/
 {\vphantom {{{\lambda _2}} {\left( {2(N - 1)} \right)}}} \right.
 \kern-\nulldelimiterspace} {\left( {2(N - 1)} \right)}}}}$ follows an $F$-distribution having degrees of freedom ${\left( 2,  {2(N - 1)} \right)}$, which is denoted by $F_{\left( 2,  {2(N - 1)} \right)}$, and the CDF of
 ${\gamma _{{E_a}}}$ is given by
\begin{align}\label{gammaEACDF}\
{F_{{\gamma _{{E_a}}}}}\left( x \right) = 1 - {\left( {\frac{{{P_A}{\sigma ^2}_{A,{E_a}}}}{{{P_A}{\sigma ^2}_{A,{E_a}} + {P_{{J_A}}}{\sigma
^2}_{J,{E_a}}x}}} \right)^{N - 1}}.
\end{align}
Similarly, according to  \eqref{signalEk}, the SNR at the $k$-th passive eavesdropper $E_k$ is expressed as \begin{align}\label{gammaEk}\
{\gamma _{{E_k}}} = \frac{{{P_A}{{\left| {{h_{A,{E_k}}}} \right|}^2}}}{{{P_{{J_A}}}{{\left| {{{\bf{g}}^H}_{{E_k}}{{{{\mbox{\small{\bf {W}}}}}}_{{E_a}}}}
\right|}^2} + \frac{{{P_{{J_p}}}}}{{N - 2}}{{\bf{g}}^H}_{{E_k}}{{\bf{W}}_{{E_p}}}{{\bf{W}}^H}_{{E_p}}{{\bf{g}}_{{E_k}}} }}.
\end{align}
Since ${\left| {{h_{A,{E_k}}}} \right|^2} \sim \exp\left( {{\sigma
^2}_{A,{E_k}}} \right)$, $\lambda _3\triangleq{\left|{{{\bf{g}}^H}_{{E_k}}{{{\mbox{\small{\bf{W}}}}}_{{E_a}}}} \right|^2} \sim Gamma\left( {1,{\sigma ^2}_{{J, E_k}}} \right)$, and \\
$\lambda _4\triangleq{{\bf{g}}^H}_{{E_k}}{{\bf{W}}_{{E_p}}}{{\bf{W}}^H}_{{E_p}}{{\bf{g}}_{{E_k}}}\sim Gamma\left( {N-2,{\sigma ^2}_{{J,E_k}}} \right)$, the CDF of the passive eavesdropper $E_k$ is given by
 \begin{align}\label{gammaEkcdf}\
{F_{{\gamma _{{E_k}}}}}\left( x \right) &= 1 - \Pr \left( {\frac{{{P_A}{{\left| {{h_{A,{E_k}}}} \right|}^2}}}{{{P_{{J_A}}}{\lambda _3} + \frac{{{P_{{J_p}}}{\lambda _4}}}{{N - 2}}}} \ge x}
 \right)\nonumber\\
 &{\overset{(a)}=} 1 -\mathbb{E}\left( {\exp \left( { - \frac{{{P_{{J_A}}}x{\lambda _3}}}{{{P_A}{\sigma ^2}_{A,{E_k}}}}} \right)} \right)\nonumber \\
 &\quad\times \mathbb{E}\left( {\exp \left( { - \frac{{{P_{{J_p}}}x{\lambda _4}}}{{\left( {N - 2} \right){P_A}{\sigma ^2}_{A,{E_k}}}}} \right)} \right)\nonumber \\
 &{\overset{(b)}=} 1 - {\left( {1 + \frac{{{P_{{J_A}}}x{\sigma ^2}_{J,{E_k}}}}{{{P_A}{\sigma ^2}_{A,{E_k}}}}} \right)^{ - 1}}\nonumber \\
 &\quad \times {\left( {1 + \frac{{{P_{{J_p}}}x{\sigma
 ^2}_{J,{E_k}}}}{{\left( {N - 2} \right){P_A}{\sigma ^2}_{A,{E_k}}}}} \right)^{2 - N}}
\end{align}
where step (a) follows from the MGF of ${{\left| {{h_{A,{E_k}}}} \right|}^2}$, and step (b) follows from the MGF of $\lambda _3$ and $\lambda _4$ \cite{Simon}. In addition, the received signal at Bob is expressed as
\begin{align}\label{signalB}\
{{y}}_{B}=\sqrt{P_A} {{h}}_{A,B} x +\sqrt{P_{E_a}} {{f}}_{E_a,B} n_{a} + {{n}}_B
\end{align}
where $x$ is the information symbol having unit power. ${n_{{a}}} \sim \mathcal {CN} \left( {0,1} \right)$ denotes the AN transmitted from $E_a$, and $P_{E_a}$ denotes the transmission power at the active eavesdropper. In addition, ${{n}}_B$ denotes the additive white Gaussian noise (AWGN) at Bob having mean zero and variance $N_0=1$. Assume that the interference power at the legitimate receiver is much greater than the noise power, i.e., an interference-limited environment for the legitimate receiver. Then, according to \eqref{signalB}, the SNR at Bob is expressed as
\begin{align}\label{gammaB}\
{\gamma _B} = \frac{{{P_A}{{\left| {{h_{A,B}}} \right|}^2}}}{{{P_{{E_a}}}{{\left| {{h_{{E_a},B}}} \right|}^2}}}.
\end{align}
Since ${\left| {{h_{A,B}}}\right|^2}\sim Gamma\left( {1,{\sigma ^2}_{A,B}} \right)$, the CDF of ${\gamma _B}$ is given by
 \begin{align}\label{gammaBcdf}\
{F_{{\gamma _B}}}\left( x \right) = 1 - \frac{1}{{{P_{{E_a}}}\sigma _{{E_a},B}^2}}{\left( {\frac{1}{{{P_{{E_a}}}\sigma _{{E_a},B}^2}} + \frac{x}{{{P_A}\sigma _{A,B}^2}}} \right)^{ - 1}}.
 \end{align}
\subsection{Optimization Problem Formulation}
For the cooperative jamming scheme, both the transmission and secrecy quality should be satisfied.  To guarantee the transmission quality, the transmission outage probability requirements should be less than a predetermined threshold $\delta$. Simultaneously, to guarantee the secrecy transmission, the SOPs for both the active eavesdropper and the passive eavesdroppers should be less than a predetermined threshold $\varepsilon$. Thus, to maximize the secrecy rate $R_s$ under reliability and security transmission requirements, we formulate the optimization problem as follows
\begin{subequations}
\begin{align}\label{problem}\
&\begin{array}{*{20}{c}}
{\mathop {\max }\limits_{\hfill\scriptstyle{P_A},{P_{{J_A}}},{P_{{J_p}}}\hfill\atop} }& R_s
\end{array}\\
&{s.t.}\quad {{p_{to}} \le \delta },\label{problemb}\\
&\quad \quad p_{so_1}\leq \varepsilon, p_{so_2}\leq \varepsilon, \label{problema}\\
&\quad\quad{P_A} + {P_{{J_A}}} + {P_{{J_p}}} \le {P_{\max }}, \label{problemc}\\
&\quad \quad  0 \leq R_s \leq R_b
\end{align}
\end{subequations}
where \eqref{problemb} and \eqref{problema} denote the reliable transmission constraint and secrecy transmission constraint, respectively. In addition, eq. \eqref{problemc} is the total power constraint. ${p_{to}}$ is the transmission outage probability and it is given by
\begin{align}\label{pto}\
{p_{to}}&= \Pr \left( {{{\log }_2}\left( {1 + {\gamma _B}} \right) < {R_b}} \right)\nonumber\\
 &= 1 - \frac{1}{{{P_{{E_a}}}\sigma _{{E_a},B}^2}}\left( {\frac{1}{{{P_{{E_a}}}\sigma _{{E_a},B}^2}} + \frac{{{2^{{R_b}}} - 1}}{{{P_A}\sigma _{A,B}^2}}} \right)^{-1}.
\end{align}
Let ${P_{{J_p}}} = \left( {1 - \theta } \right)\left( {{P_{\max }} - {P_A}} \right)$ and ${P_{{J_A}}} = \theta \left( {{P_{\max }} - {P_A}} \right)$ where  $0\leq \theta \leq 1$. Then, according to \eqref{gammaEACDF}, the SOP for the active eavesdropper $E_a$, is expressed as
 \begin{align}\label{pso1}\
{p_{so_1}} &= \Pr \left( {{{\log }_2}\left( {1 + {\gamma _{{E_a}}}} \right) \ge {R_b} - {R_s}} \right)\nonumber \\
  &= {\left( {1 + \theta\alpha} \right)^{1 - N}}
\end{align}
where $\alpha=\left( {\frac{{{P_{\max }}}}{{{P_A}}} - 1} \right)\frac{{\sigma _{J,E_a}^2\left( {{2^{{R_b} - {R_s}}} - 1} \right)}}{{\sigma _{A,E_a}^2}}$ is greater than zero since ${\frac{{{P_{\max }}}}{{{P_A}}}}\leq 1$ and ${{2^{{R_b} - {R_s}}} - 1}\geq 0$. In addition,  since it is challenging for the  passive eavesdroppers to acquire the perfect instantaneous CSIs of the jammer $\rightarrow$ $E_k$  links, the eavesdropper that has the maximum SNR overhears the legitimate message at each time slot. Then, according to \eqref{gammaEkcdf}, the SOP for the passive eavesdroppers is given by
 \begin{align}\label{pso2}\
{p_{so_2}} &= \Pr \left( {\mathop {\max}\limits_{1 \le k \le K} \left( {{{\log }_2}\left( {1 + {\gamma _{{E_k}}}} \right)} \right) \ge {R_b} - {R_s}} \right)\nonumber \\
 &= 1 - {\left( {1 - \frac{1}{{1 + \beta \theta }}{{\left( {1 + \frac{{\beta \left( {1 - \theta } \right)}}{{\left( {N - 2} \right)}}} \right)}^{2 - N}}} \right)^K}
\end{align}
where $\beta  = \left( {\frac{{{P_{\max }}}}{{{P_A}}} - 1} \right)\frac{{\left( {{2^{{R_b} - {R_s}}} - 1} \right)\sigma _{J,{E_k}}^2}}{{\sigma _{A,{E_k}}^2}}$. Since $\frac{{\partial {p_{so_1}}}}{{\partial {P_A}}} > 0$ and $\frac{{\partial {p_{so_2}}}}{{\partial {P_A}}} > 0$,  both $p_{so_1}$ and $p_{so_2}$ increase with $P_A$.
Also, since $\frac{{\partial {p_{so_1}}}}{{\partial {R_s}}} > 0$ and $\frac{{\partial {p_{so_2}}}}{{\partial {R_s}}} > 0$, both
$p_{so_1}$ and $p_{so_2}$ increase with $R_s$.
\subsection{Parameter Optimization for Secrecy Rate Maximization}
 It can be shown that both $p_{so_1}$ and $p_{so_2}$ are monotonous increasing functions with $P_A$ and $R_s$. Thus, to maximize $R_s$, we should first derive the minimum required $P_A$. According to \eqref{problemb} and \eqref{pto}, the minimum power allocated to Alice is given by
 \begin{align}\label{pa}\
 {P^*_A} = \frac{{1 - {2^{{R_b}}}}}{{\ln \left( {1 - \delta } \right){\sigma ^2}_{A,B}}}.
\end{align}
It is clear that if ${P^*_A}> P_{\max}$, there is no solution to \eqref{problem}. After determining the optimal power $P^*_A$,  we can rewrite the optimization problem in \eqref{problem} as
\begin{align}\label{problem1}\
&\begin{array}{*{20}{c}}
{\mathop {\max }\limits_{\hfill\scriptstyle \theta\hfill} }&R_{s}
\end{array}\nonumber \\
{s.t.}&\quad p_{so_1}\leq \varepsilon, p_{so_2}\leq \varepsilon, \nonumber \\
&\quad{ 0 \leq \theta  \leq 1 },\nonumber \\
&\quad  0 \leq R_s \leq R_b.
\end{align}
In \eqref{problem1}, the first-order derivative of $p_{so_1}$ with respect to (w.r.t.) $\theta$ is given by
\begin{align}\label{pso1derivation}\
\frac{{\partial {p_{so_1}}}}{{\partial \theta }} = \left( {1 - N} \right)\alpha{\left( {1 + \theta\alpha} \right)^{ - N}}.
\end{align}
  Obviously $1 - N\le 0$ and $\frac{{\partial {p_{so_1}}}}{{\partial \theta }}< 0$. This means that there is a monotonous relationship between $p_{so_1}$ and $\theta$, i.e., $p_{so_1}$ decreases with $\theta$. Then, according to the SOP requirement at the active eavesdropper, $\theta$ should satisfy the following constraint
\begin{align}\label{constraint1}\
\psi \left( {{P_A},{R_s}} \right)\leq \theta \leq 1
\end{align}
where $\psi \left( {{P_A},{R_s}} \right) =\alpha^{-1} \left( {{\varepsilon ^{\frac{1}{{1 - N}}}} - 1} \right)$. Note that if $\psi  \left( {{P_A},{R_s}} \right)> 1$, there is no solution to \eqref{problem}. In addition, the first-order derivative of $p_{so_2}$ w.r.t. $\theta$ is derived as
\begin{align}\label{pso2derivation}\
\frac{{\partial {p_{so_2}}}}{{\partial \theta }} &= K{\left[ {1 - \frac{1}{{1 + \beta \theta }}{{\left( {1 + \frac{{\beta \left( {1 - \theta } \right)}}{{{N - 2}}}} \right)}^{2- N}}} \right]^{K - 1}}\nonumber \\
 &\times {\left( {\frac{\beta }{{1 + \beta \theta }}} \right)^2}{\left( {1 + \frac{{\beta \left( {1 - \theta } \right)}}{{{N - 2}}}} \right)^{ 1- N}}\left( {\frac{{N\theta -\theta - 1}}{{N - 2}}} \right).
\end{align}
It can be shown that $\frac{{\partial {p_{so_2}}}}{{\partial \theta }} <0$ when  $0\leq \theta <\frac{1}{N-1}$, $\frac{{\partial {p_{so_2}}}}{{\partial \theta }} =0$ when $\theta =\frac{1}{N-1}$,  and $\frac{{\partial {p_{so_2}}}}{{\partial \theta }} >0$ when  $\frac{1}{N-1}< \theta \leq 1$. Thus, for a fixed secrecy rate $R_s$ and a fixed transmission power $P_A$, $p_{so_2}$ is a convex function of $\theta$, and $p_{so_2}$ is minimized when $\theta=\frac{1}{N-1}$. Assume ${\varphi \left( {{P_A},{R_s}}\right)}$ is the inverse function of $p_{so_2}$. Since $p_{so_2}$ is a convex function of $\theta$, there are two $\theta$ values satisfying $p_{so_2}=\varepsilon$, namely, the smaller value $\min {\left( {\varphi \left( {{P_A},{R_s}} \right)} \right)}$ and the larger value $\max {\left( {\varphi\left({{P_A},{R_s}} \right)} \right)}$. When $\varepsilon$ is large, $\min {\left( {\varphi \left( {{P_A},{R_s}} \right)} \right)}$ can be less than zero and  $\max {\left( {\varphi \left( {{P_A},{R_s}} \right)} \right)}$ can be larger than unity. However, $\theta$ has been defined as the ratio of $P_{J_A}$ to $P_{\max}-P_A$ and has a range between zero and unity. Thus, to satisfy $p_{so_2}\leq\varepsilon$, $\theta$ is given by
\begin{align}\label{constraint2}\
\max\left(0, \min \left( {\varphi \left( {{P_A},{R_s}} \right)} \right)\right) < \theta <
 \min \left(1, \max \left( {{\varphi}\left( {{P_A},{R_s}} \right)} \right)\right).
\end{align}
Note that if $p_{so_2}>\varepsilon$ when $\theta=\frac{1}{N-1}$, Eq. \eqref{constraint2} is satisfied and there is no solution to \eqref{problem}.

According to \eqref{pa}, \eqref{constraint1}, and \eqref{constraint2}, the maximum secrecy rate can be obtained by Algorithm 1. In this algorithm, $\Delta$ denotes the incremental step for $R_s$, and it can be determined by the required $R_s$ accuracy. In addition, we have the following remarks regarding Algorithm 1.

\begin{algorithm}[h]
\caption{Secrecy Rate $R_s$ Maximization}
\begin{algorithmic}[1]
\STATE {Initialization: $N, K$, $\varepsilon$, $P_{\max}$, $R_s\in [0, R_b)$}.
\STATE {According to \eqref{pa}, the optimal power allocated to Alice $P^*_A$ can be obtained}.
\STATE {Set $i=1$, $R_s=0$},
\STATE Calculate the constraint \eqref{constraint1} and \eqref{constraint2}\\
     if $\psi \left( {{P^*_A},{R_s}} \right)< \min \left(1, \max \left( {\varphi \left( {{P^*_A},{R_s}} \right)} \right)\right)$, $R_s=R_s+\Delta$,\\
     else break;
\STATE  {$i=i+1$, repeat step 4 until $\psi \left( {{P^*_A},{R_s}} \right) \geq \min \left(1, \max \left( {\varphi \left( {{P^*_A},{R_s}} \right)}, \right)\right)$,\\
or there is no solution for the constraints in \eqref{problem1}}.
\STATE  {Then $\theta^*=\psi \left( {{P^*_A},{R_s}} \right)$, the maximum secrecy rate $R^*_s= R_s$}.
\end{algorithmic}
\end{algorithm}

   {\emph{Remark 1:}} For the proposed scheme having a fixed total transmit power, the maximum secrecy rate $R_s$ decreases with $R_b$, because large $R_b$ requires more power to be allocated at Alice and the remaining power for the AN is decreased.

   {\emph{Remark 2:}} Under the secrecy outage constraint, the AN power for the active eavesdropper is inversely proportional to the channel quality between the jammer and the active eavesdropper, and is inversely proportional to the number of transmit antennas. For the extreme case when the active eavesdropper is close to the jammer or the number of transmit antennas $N$ approaches infinity, the AN power for the active eavesdropper goes to zero.

  {\emph{Remark 3:}} Under the secrecy outage constraint, there is no monotonous relationship between the AN power for the passive eavesdroppers and $\sigma^2_{J,E_k}$. Specifically, when the passive eavesdropper is close to the jammer, $\theta^*$ approaches unity. It is noted that the optimal $\theta^*$  corresponding to the maximum secrecy rate is equal to $1/(N-1)$, which also corresponds to the minimum value of SOP for the passive eavesdroppers. This means that the SOP for the passive eavesdroppers dominates the maximum secrecy rate and the SOP for the active eavesdropper is negligible.

In addition, since $p_{so_1}$ is a monotonous function of both $R_s$ and $\theta$, Algorithm $1$ stops when $\theta=1$ cannot satisfy the $p_{so_1}$ constraint. By contrast, since $p_{so_2}$ is a convex function of $\theta$, Algorithm $1$ stops only when there is no solution for $\theta$ that satisfies SOP constraints.

\section{Optimal Power Allocation With Imperfect CSI}
In our proposed jamming scheme, the beamforming vector depends on the CSI of the jammer to Bob and the active eavesdropper links. However, it is challenging to acquire the perfect CSI due to feedback delay or estimation error. Therefore, we will investigate the impact of imperfect ${\bf{g}}_B $ and imperfect ${\bf{g}}_{E_a}$ on the maximum secrecy rate in the following subsections.
\subsection{Imperfect CSI Between Jammer and Bob}
Assume that the instantaneous CSI ${{\bf{g}}_B}$ cannot be perfectly estimated at the jammer, and ${{\tilde{\bf{g}}}_{B}} \sim \mathcal {CN} \left( {{\bf{0}},\sigma^2_{J,B}{{\bf{I}}_{N}}} \right)$ denotes the imperfect CSI between the jammer and Bob. The relationship between ${{\bf{g}}_B}$ and ${{\tilde{\bf{g}}}_{B}}$ is expressed as \cite{JKim}
\begin{align}\label{imperfectgB}\
{{\bf{g}}_B} = {\rho _B}{\tilde{{\bf{g}}}_B} + {{\bf{e}}_B}, \quad\quad 0 \leq {{{\rho}}_{{B}}}\leq 1
\end{align}
where ${{{\bf{e}}_B}} \sim \mathcal {CN} \left( {{\bf{0}},\left( 1-{\rho^2 _B}\right)\sigma^2_{J,B}{{\bf{I}}_{N}}} \right)$ denotes the estimation error for ${\bf{g}}_{B}$, and $\rho _B$ is the correlation coefficient between ${\bf{g}_B} $ and ${\tilde{\bf{g}}_B}$. Since ${\tilde{\bf{g}}_B}$ only determines the null-space of ${{{{\mbox{\small{\bf {W}}}}}}_{{E_a}}}$ and ${\bf{W}}_{{E_p}}$, it will not affect the SNR at both the active eavesdropper and the passive eavesdroppers, which have the same expressions as \eqref{gammaEA} and \eqref{gammaEk}, respectively. However due to imperfect CSI ${\tilde{\bf{g}}_B}$, the AN generated to interfere with the eavesdroppers will leak to Bob, and the SNR at Bob is modified to
\begin{align}\label{imperfectgB}\
{\tilde \gamma _B} = \frac{{{P_A}{{\left| {{h_{A,B}}} \right|}^2}}}{{{P_{{J_A}}}{{\left| {{{{{\tilde {\mbox{\small{{\bf{W}}}}}}}}^H}_{{E_a}}{{\bf{e}}_B}} \right|}^2} + \frac{{{P_{{J_p}}}}}{{N - 2}}{\left\| {{{{\bf{\tilde W}}}^H}_{{E_p}}{{\bf{e}}_B}} \right\|}^2 }}
\end{align}
where ${{{\left| {{{{{\tilde{\mbox{\small{{\bf {W}}}}}}}}^H}_{{E_a}}{{\bf{e}}_B}} \right|}^2}}$ follows an exponential distribution having parameter ${\left( {1 - {\rho ^2}_B} \right){\sigma ^2}_{J,B}} $. The ratio $\frac{{2\left\| {{{{\bf{\tilde W}}}^H}_{{E_p}}{{\bf{e}}_B}} \right\|}}{{\left( {1 - {\rho ^2}_B} \right){\sigma ^2}_{{J,B}}}}$ follows a chi-square distribution having \emph{n} degrees of freedom $\chi^2(2(N-2))$. \\
Thus, ${{{P_{{J_A}}}{{\left| {{{{{\tilde {\mbox{\small{{\bf {W}}}}}}}}^H}_{{E_a}}{{\bf{e}}_B}} \right|}^2} + \frac{{{P_{{J_p}}}}}{{N - 2}}{\left\| {{{{\bf{\tilde W}}}^H}_{{E_p}}{{\bf{e}}_B}} \right\|}^2 }}\leq \left( {{P_{{J_A}}} + {P_{{J_P}}}} \right)\left( {1 - {\rho ^2}_B} \right){\sigma ^2}_{J,B}$ \cite{XZhou1,TZheng}, and an upper bound on $ {\tilde\gamma _B}$ is expressed as
\begin{align}\label{imperfectgB}\
{\tilde\gamma _B} \le \frac{{{P_A}{{\left| {{h_{A,B}}} \right|}^2}}}{{\left( {{P_{\max }} - {P_A}} \right)\left( {1 - {\rho ^2}_B} \right){\sigma ^2}_{J,B}}}.
\end{align}
Using \eqref{imperfectgB}, we obtain the transmission outage probability as follows
\begin{align}\label{pto2}\
&{P_{to}}= \left( {\frac{{{{\left| {{h_{A,B}}} \right|}^2}}}{{\left( {\frac{{{P_{\max }}}}{{{P_A}}} - 1} \right)\left( {1 - {\rho ^2}_B} \right){\sigma ^2}_{J,B}}} \le {2^{{R_b}}} - 1} \right)\nonumber \\
&  = 1 - \exp \left( { - \frac{{\left( {\frac{{{P_{\max }}}}{{{P_A}}} - 1} \right)\left( {1 - {\rho ^2}_B} \right){\sigma ^2}_{J,B}\left( {{2^{{R_b}}} - 1} \right)}}{{{\sigma ^2}_{A,B}}}} \right).
\end{align}
Substituting \eqref{pto2} into \eqref{problemb}, we derive the required transmission power at Alice as
\begin{align}\label{pto22}\
{P_A} \ge \frac{{{P_{\max }}}}{{1 - \frac{{{\sigma ^2}_{A,B}\ln \left( {1 - \delta } \right)}}{{\left( {1 - {\rho ^2}_B} \right){\sigma ^2}_{J,B}\left( {{2^{{R_b}}} - 1} \right)}}}}
\end{align}
where $0 <{{{\rho}}_{{B}}}< 1$. From \eqref{pto22}, we can obtain the minimum required power at Alice as ${P^*_A} = \frac{{{P_{\max }}}}{{1 - \frac{{{\sigma ^2}_{A,B}\ln \left( {1 - \delta } \right)}}{{\left( {1 - {\rho ^2}_B} \right){\sigma ^2}_{J,B}\left( {{2^{{R_b}}} - 1} \right)}}}}$. Once $P^*_A$ is obtained, we can still use Algorithm 1 to obtain the maximum secrecy rate $R^*_s$. This is due to the fact that the SOPs for the active and passive eavesdroppers have the same expressions as \eqref{pso1} and \eqref{pso2}, respectively. Also, when ${{{\rho}}_{{B}}}=1$, the optimal parameters, $\theta^*$ and $P^*_A$, have the same values as in the perfect CSI case. When ${{{\rho}}_{{B}}}=0$, the jammer cannot obtain the instantaneous CSI between the jammer and Bob, which means that more AN will leak to Bob, and more power is required at Alice to satisfy \eqref{problemb}.
\subsection{Imperfect CSI Between Jammer and The Active Eavesdropper}
In practice, the instantaneous CSI between the jammer and the active eavesdropper can also be imperfect, which we assume to be described as ${{\tilde{\bf{g}}}_{E_a}} \sim \mathcal {CN} \left( {{\bf{0}},\sigma^2_{J,E_a}{{\bf{I}}_{N}}} \right)$. The relationship between ${{\bf{g}}_{{E_a}}}$ and ${{\tilde{ \bf{g}}}_{{E_a}}}$ is given by
\begin{align}\label{imperfectEA}\
{{\bf{g}}_{{E_a}}} = {{{\rho}}_{{E_a}}}{{\tilde{ \bf{g}}}_{{E_a}}} + {{\bf{e}}_{{E_a}}}, \quad\quad\quad  0 \leq {{{\rho}}_{{E_a}}} \leq 1
\end{align}
where ${{\bf{e}}_{E_a}} \sim \mathcal {CN} \left( {{\bf{0}},\left( 1-{\rho^2 _{E_a}}\right)\sigma^2_{J,E_a}{{\bf{I}}_{N}}} \right)$ denotes the estimation error for ${\bf{g}}_{E_a}$, and $\rho _{E_a}$ is the correlation coefficient between ${{\tilde{\bf{g}}}_{E_a}} $ and ${{\bf{g}}_{E_a}}$.  Since ${{{{{\tilde{\mbox{\small \bf {W}}}}}}_{{E_a}}}} = \frac{{\left( {{{\bf{I}}_N} - \frac{{{{\bf{g}}_B}{{\bf{g}}^H}_B}}{{{{\left\| {{{\bf{g}}_B}} \right\|}^2}}}} \right){\tilde{\bf{g}}^*}_{{E_a}}}}{{\left\| {\left( {{{\bf{I}}_N} - \frac{{{{\bf{g}}_B}{{\bf{g}}^H}_B}}{{{{\left\| {{{\bf{g}}_B}} \right\|}^2}}}} \right){\tilde{\bf{g}}^*}_{{E_a}}} \right\|}}$ lies in the null space of ${{\bf{g}}_B}$, ${\tilde{\bf{g}}_{E_a}}$ does not affect the SNR at Bob. At the same time,  the SNR at the \emph{k}th passive eavesdropper is expressed as
\begin{align}\label{gammaEk1}\
{\tilde \gamma _{{E_k}}} = \frac{{{P_A}{{\left| {{h_{A,{E_k}}}} \right|}^2}}}{{{P_{{J_A}}}{{\left| {{{\bf{g}}^H}_{{E_k}}{{{{\tilde{\mbox{\small{\bf {W}}}}}}}_{{E_a}}}}
\right|}^2} + \frac{{{P_{{J_p}}}}}{{N - 2}}{{\bf{g}}^H}_{{E_k}}{{\bf{\tilde{W}}}_{{E_p}}}{{\bf{\tilde{W}}}^H}_{{E_p}}{{\bf{g}}_{{E_k}}} }}
\end{align}
where  ${{{\bf{\tilde {W}}}}_{{E_p}}}\in \mathbb{C}^{N\times(N-2)}$ lies in the null-space of $\left[{{\bf{g}}_B},{{{{{\tilde {\mbox{\small \bf {W}}}}}}_{{E_a}}}}\right]$.  Since ${\left| {{h_{A,{E_k}}}} \right|^2} \sim Gamma\left( {1,{\sigma
^2}_{A,{E_k}}} \right)$, ${\left|{{{\bf{g}}^H}_{{E_k}}{{{\tilde{\mbox{\small{\bf {W}}}}}}_{{E_a}}}} \right|^2} \sim Gamma\left( {1,{\sigma ^2}_{{J, E_k}}} \right)$, and ${{\bf{g}}^H}_{{E_k}}{{\bf{\tilde{W}}}_{{E_p}}}{{\bf{\tilde{W}}}^H}_{{E_p}}{{\bf{g}}_{{E_k}}}\sim Gamma\left( {N-2,{\sigma ^2}_{{J,E_k}}} \right)$, the CDF of $\tilde \gamma _{{E_k}}$ has the same expression as \eqref{gammaEkcdf}. Thus, the SOP for the passive eavesdroppers $p_{so_2}$ has the same expression as \eqref{pso2}. Though ${{\tilde{ \bf{g}}}_{{E_a}}}$ does not change the SNRs at Bob and the passive eavesdroppers, it affects the SNR at the active eavesdropper. The SOP for the active eavesdropper is presented in the following lemma.

{\emph{Lemma 1:}} With ${{\tilde {\bf{g}}}_{{E_a}}}$, the SOP for the active eavesdropper is given by
 \begin{align}\label{pso1imperfect}\
 p_{so_1}&=\frac{{{{\left( {1 + \theta \left( {1 - {\rho ^2}_{{E_a}}} \right)\alpha } \right)}^{N - 2}}}}{{{{\left( {1 + \theta \alpha } \right)}^{N - 1}}}}\nonumber \\
 &\times {\left( {1 + \frac{{\left( {1 - \theta } \right)\left( {1 - \rho _{{E_a}}^2} \right)\alpha }}{{N - 2}}} \right)^{2 - N}}.
 \end{align}
 {\emph{Proof:}} The proof is presented in Appendix A. When $\rho _{{E_a}}=1$, Eq. \eqref{pso1imperfect} specializes to \eqref{pso1}. After obtaining the expression of $p_{so_1}$, we quantify the impact of $R_s$, $P_A$ and $\theta$ on $p_{so_1}$ in the following lemma

{\emph{Lemma 2:}} With fixed  $P_A$, $R_s$, and $\theta$, $p_{so_1}$ decreases with $\rho_{E_a}$, while with fixed  $P_A$ and $\theta$, $p_{so_1}$ increases with $R_s$. In addition, with fixed $R_s$ and $\theta$, $p_{so_1}$ increases with $P_A$. With fixed $P_A$ and $R_s$, when $\theta_2$, the positive solution of $\frac{{\partial {P_{so_1}}}}{{\partial \theta }} = 0 $, is larger than unity, $\frac{{\partial {P_{so_1}}}}{{\partial \theta }} < 0 $. Otherwise, when $\theta_2 \leq 1$, $\frac{{\partial {P_{so_1}}}}{{\partial \theta }} > 0 \left(0\leq\theta\leq\theta_2 \right)$ and $\frac{{\partial {P_{so_1}}}}{{\partial \theta }} < 0 \left(\theta_2\leq\theta\leq 1\right)$.

{\emph{Proof:}} The proof is presented in Appendix B. Note that when $p_{so_1}=\varepsilon$, we can obtain $\theta=\varsigma \left( {{R_s},{P_A}} \right)$, where $\varsigma \left( {{R_s},{P_A}} \right)$ is the inverse function of $p_{so_1}$. Though $\varsigma \left( {{R_s},{P_A}} \right)$ can be less than zero or greater than unity, a reasonable value of $\theta$ should lie between zero and unity. Then according to Lemma 2, to satisfy the constraint $p_{so_1}\leq\varepsilon$, two cases arise according to \eqref{theta2large1}.

\emph{Case 1:} Eq. \eqref{theta2large1} is satisfied, which means that $p_{so_1}$ decreases with $\theta$ $\left(0 \le \theta  \le 1\right)$. Then, to satisfy the requirement of $p_{so_1}\leq\varepsilon$, the range of $\theta$ is
\begin{align}\label{constraint3}\
\max \left( {0,\varsigma \left( {{R_s},{P_A}} \right)} \right) \le \theta  \le 1
\end{align}
where $\varsigma \left( {{R_s},{P_A}} \right)\leq 1$ is required.

\emph{Case 2:} Eq. \eqref{theta2large1} is not satisfied. Then, $p_{so_1}\leq\varepsilon$ is a convex function of $\theta$, where  $p_{so_1}$ decreases with $\theta$  $\left(0 \le \theta  \le \theta_2\right)$ and increases with  $\theta$  $\left(\theta_2 \le \theta  \le 1\right)$.  Thus, if $p_{so_1}=\varepsilon$, there are two solutions for $\theta$, namely, $\min\left( \varsigma \left( {{R_s},{P_A}} \right)\right) $ and $\max\left( \varsigma \left( {{R_s},{P_A}} \right)\right)$. To satisfy the requirement of $p_{so_1}\leq\varepsilon$,  the range of $\theta$ is
\begin{align}\label{constraint4}\
 \max \left(0, \min\left( \varsigma \left( {{R_s},{P_A}} \right)\right)\right) \leq\theta\leq  \min \left(1, \max\left( \varsigma \left( {{R_s},{P_A}} \right)\right)\right).
\end{align}
 For fixed $R_s$ and $P_A$, if there is no $\theta$ that satisfies \eqref{constraint3} or \eqref{constraint4}, no solution exists for \eqref{problem}. Thus, according to Lemma 2, \eqref{constraint3}, and \eqref{constraint4}, Algorithm 2 can be used to obtain the maximum secrecy rate. In Algorithm 2, $R_s$ increases until there is no $\theta$ available to satisfy the SOP constraints at both the active and passive eavesdroppers, or the secrecy rate $R_s$ is greater than the transmission rate $R_b$.

\begin{algorithm}[h]
\caption{Secrecy Rate $R_s$ Maximization With ${{\tilde{ \bf{g}}}_{{E_a}}}$}
\begin{algorithmic}[1]
\STATE {Initialization: $N, K$, $\varepsilon$, $P_{\max}$, $R_s\in [0, R_b)$}.
\STATE {According to \eqref{pa}, the optimal power allocated to Alice $P^*_A$ can be obtained}.
\STATE {Set $i=1$, $R_s=0$},
\STATE  {if \eqref{theta2large1} is satisfied\\
           \quad {if \eqref{constraint2} and \eqref{constraint3} have no intersecting values,}
                {$R_s=R_s+\Delta$},\\
           \quad else   break;\\
        else if \eqref{theta2large1} is not satisfied\\
            \quad {if \eqref{constraint2} and \eqref{constraint4} have no intersecting values,}
                {$R_s=R_s+\Delta$}, \\
           \quad else   break;
        }
\STATE  {$i=i+1$, repeat step 4 until that $R_s\geq R_b$.}
\STATE  {Then the maximum secrecy rate $R^*_s= R_s$}.
\end{algorithmic}
\end{algorithm}
{\emph{Remark 4:}} Different from the perfect ${{\bf{g}}_{{E_a}}}$ case in Section III, $p_{so_1}$ can be a convex function of $\theta$ when \eqref{theta2large1} is not satisfied. This is because under the case of imperfect CSI ${{\tilde{ \bf{g}}}_{{E_a}}}$, the AN injected towards the passive eavesdroppers leaks to the active eavesdropper. Of course, similar to the case of perfect ${{\bf{g}}_{{E_a}}}$,  $p_{so_1}$  decreases with $\theta$ when $N$ goes to infinity or $\sigma^2_{J,E_a}/\sigma^2_{A,E_a}$ goes to infinity. In this case, Eq. \eqref{theta2large1} always holds.

{\emph{Remark 5:}} Different from the perfect ${{\bf{g}}_{{E_a}}}$ case in Section III, $\theta^*$ can be equal to zero with imperfect CSI ${{\tilde{ \bf{g}}}_{{E_a}}}$. The AN injected towards the passive eavesdroppers can leak to the active eavesdropper due to imperfect CSI ${{\tilde{ \bf{g}}}_{{E_a}}}$. For this case, the SOP constraint $\varepsilon$ may be small and $\sigma^2_{J,E_a}$ is large. In addition, when the SOP for the active eavesdropper is far less than the SOP constraint $\varepsilon$, the SOP for the passive eavesdroppers determines the secrecy rate and  $\theta^*=1/(N-1)$, which has no relation with the correlation coefficient $\rho_{E_a}$. On the other hand, when the SOP for the passive eavesdroppers is far less than the SOP constraint $\varepsilon$, the SOP for the active eavesdropper determines the secrecy rate, and $\theta^*=\theta_2$, which has no relationship to the correlation coefficient $\rho_{E_a}$.
\section{Multiple Active Eavesdroppers}
The proposed jamming scheme can be extended to the multiple active eavesdroppers case when the number of antennas at the jammer is greater than the number of active eavesdroppers. In the presence of $M$ active eavesdroppers and $N$ passive eavesdroppers available, it is challenging to design the optimal beamforming vector towards the active eavesdroppers. Hence, for the two-fold zero-forcing scheme, MRT is still used for secrecy transmission.  Thus, the beamforming vector towards the active eavesdropper is ${{{{{\mbox{\bf{V}}}}}_{{E_{a}}}}}=\left[{{{{{\mbox{\small\bf{W}}}}}_{{E_{a_1}}}}},\cdot\cdot\cdot,{{{{{\mbox{\small\bf{W}}}}}_{{E_{a_m}}}}},\cdot\cdot\cdot,{{{{{\mbox{\small\bf{W}}}}}_{{E_{a_M}}}}}\right]$ $\left(1 \leq m \leq M \right )$, where ${{{{{\mbox{\small\bf{W}}}}}_{{E_{a_m}}}}}\in \mathbb{C}^{N\times 1}$ denotes the beamforming vector towards the \emph{m}th active eavesdroppers, and it is written as
\begin{align}\label{27}\
{{{{{\mbox{\small\bf{W}}}}}_{{E_{a_m}}}}} = \frac{{\left( {{{\bf{I}}_N} - \frac{{{{\bf{g}}_B}{{\bf{g}}^H}_B}}{{{{\left\| {{{\bf{g}}_B}} \right\|}^2}}}} \right){{\bf{g}}^*}_{{E_{a_m}}}}}{{\left\| {\left( {{{\bf{I}}_N} - \frac{{{{\bf{g}}_B}{{\bf{g}}^H}_B}}{{{{\left\| {{{\bf{g}}_B}} \right\|}^2}}}} \right){{\bf{g}}^*}_{{E_{a_m}}}} \right\|}}.
\end{align}
For analytical tractability, assume that all the active eavesdroppers have the same statistical CSI, i.e., $\sigma^2_{J,E_{a_m}}= \sigma^2_{J,E_{a}}$. Then, the AN power $P_{J_A}$ for the active eavesdroppers is equally allocated to the directions of all active eavesdroppers. In addition, since the AN towards the $m'$th active eavesdropper can leak to the $m$th passive eavesdropper, different from the single active eavesdropper case, the SNR at the $m$th active eavesdropper is given by
\begin{align}\label{27}\
{\gamma _{{E_{{a_m}}}}} = \frac{{{P_A}{{\left| {{h_{A,{E_{{a_m}}}}}} \right|}^2}}}{{\frac{{{P_{{J_A}}}}}{M}{{\left| {{{\bf{g}}^H}_{{E_{{a_m}}}}{{{\mbox{\small\bf{W}}}}_{{E_{{a_m}}}}}} \right|}^2} + \frac{{{P_{{J_A}}}}}{M}\sum\limits_{m' = 1,m' \ne m}^M {{{\left| {{{\bf{g}}^H}_{{E_{{a_{m'}}}}}{{{\mbox{\small\bf{W}}}}_{{E_{{a_m}}}}}} \right|}^2}} }}
\end{align}
where ${\lambda _7} \triangleq\frac{{{{\left| {{{\bf{g}}^H}_{{E_{{a_m}}}}{{\bf{W}}_{{E_{{a_m}}}}}} \right|}^2}}}{{{\sigma ^2}_{J,{E_a}}/2}}\sim \chi^2(2(N-1))$, and ${\lambda _8} \triangleq \frac{{\sum\limits_{m' = 1,m' \ne m}^M {{{\left| {{{\bf{g}}^H}_{{E_{{a_{m'}}}}}{{\bf{W}}_{{E_{{a_m}}}}}} \right|}^2}} }}{{{\sigma ^2}_{J,{E_a}}/2}}\sim \chi^2(2(M-1))$. Then, ${\lambda _7}{\rm{ + }}{\lambda _8} \sim \chi^2(2(N+M-2))$.  Thus, the ratio $\frac{{{{{\lambda _1}}
\mathord{\left/
 {\vphantom {{{\lambda _1}} 2}} \right.
 \kern-\nulldelimiterspace} 2}}}{{{{{\left( {\lambda _7}+{\lambda _8}\right)}} \mathord{\left/
 {\vphantom {{{{\lambda _7}+{\lambda _8}}} {\left( {2(N - 1)} \right)}}} \right.
 \kern-\nulldelimiterspace} {\left( {2(N+M - 2)} \right)}}}}$ follows an $F$-distribution having $\left( 2,  {2(N+M-2)} \right)$ degrees of freedom, which is denoted by $F_{\left( 2,  {2(N+M-2)} \right)}$, and the CDF of
 ${\gamma _{{E_{a_m}}}}$ is given by
\begin{align}\label{gammaEACDF1}\
{F_{{\gamma _{{E_{a_m}}}}}}\left( x \right) = 1 - {\left( {\frac{{{P_A}{\sigma ^2}_{A,{E_a}}}}{{{P_A}{\sigma ^2}_{A,{E_a}} + \frac{{P_{{J_A}}}}{M}{\sigma
^2}_{J,{E_a}}x}}} \right)^{M+N - 2}}.
\end{align}
According to \eqref{gammaEACDF1}, the secrecy outage probability for the active eavesdroppers is given by
\begin{align}\label{pso11}\
{p_{{{so}_1}}} = 1 - \prod\limits_{m = 1}^M {{F_{{\gamma _{{E_{{a_m}}}}}}}\left( {{2^{{R_b} - {R_s}}} - 1} \right)}
\end{align}
where selection combining is applied at the active eavesdroppers. In addition, ${{{\bf{W}}}_{{E_p}}}\in \mathbb{C}^{N\times\left( N-M-1\right)}$ should lie in the null-space of $\left[{{\bf{g}}_B},{{{{\bf{\mbox{\bf{W}}}}}_{{E_a}}}}\right]$. Similar to the single active eavesdropper case, the AN for the active eavesdroppers will leak to the passive eavesdroppers. Then, the SNR at the $k$th passive eavesdropper is rewritten as
\begin{align}\label{gammaEkCDF}\
{\gamma _{{E_k}}} = \frac{{{P_A}{{\left| {{h_{A,{E_k}}}} \right|}^2}}}{{\frac{{{P_{{J_A}}}}}{M}\sum\limits_{m = 1}^M {{{\left| {{{\bf{g}}^H}_{{E_k}}{{{\mbox{\small\bf{W}}}}_{{E_{{a_m}}}}}} \right|}^2} + \frac{{{P_{{J_{\rm{p}}}}}}}{{N - M - 1}}{{\bf{g}}^H}_{{E_k}}{{\bf{W}}_{{E_p}}}{{\bf{W}}^H}_{{E_p}}{{\bf{g}}_{{E_k}}}} }}
\end{align}
where $\lambda_9 \triangleq {{\sum\limits_{m = 1}^M {{{\left| {{{\bf{g}}^H}_{{E_k}}{{{\mbox{\small\bf{W}}}}_{{E_{{a_m}}}}}} \right|}^2}} }} \sim Gamma (M, \sigma^2_{J,{E_k}})$. ${\lambda _{10}} \triangleq \frac{{{{\bf{g}}^H}_{{E_k}}{{\bf{W}}_{{E_p}}}{{\bf{W}}^H}_{{E_p}}{{\bf{g}}_{{E_k}}}}}{{{\sigma ^2}_{J,{E_k}}/2}} \sim\\
Gamma\left( {N-M-1,{\sigma ^2}_{{J, E_k}}} \right)$. Similar to \eqref{gammaEkcdf}, the CDF of ${\gamma _{{E_k}}}$ can be rewritten as
 \begin{align}\label{gammaEkcdf1}\
 {F_{{\gamma _{{E_k}}}}}\left( x \right) &= 1 - \Pr \left( {\frac{{{P_A}{{\left| {{h_{A,{E_k}}}} \right|}^2}}}{{\frac{P_{{J_A{\lambda_9}}}}{M}{} + \frac{{{P_{{J_p}}}{\lambda_{10}}}}{{N -M-1}}}} \ge x}
 \right)\nonumber\\
 &{=} 1 - {\left( {1 + \frac{{{P_{{J_A}}}x{\sigma ^2}_{J,{E_k}}}}{{M{P_A}{\sigma ^2}_{A,{E_k}}}}} \right)^{ - M}}\nonumber \\
 &\quad \times {\left( {1 + \frac{{{P_{{J_p}}}x{\sigma
 ^2}_{J,{E_k}}}}{{\left( {N-M-1} \right){P_A}{\sigma ^2}_{A,{E_k}}}}} \right)^{1+M - N}}
 \end{align}
Then, the secrecy outage probability for the passive eavesdroppers is rewritten as
\begin{align}\label{pso22}\
{p_{{{so}_2}}} = 1 - \prod\limits_{k = 1}^K {{F_{{\gamma _{{E_k}}}}}\left( {{2^{{R_b} - {R_s}}} - 1} \right)}.
\end{align}
Substituting \eqref{pso11} and \eqref{pso22} into \eqref{problema}, the optimal power allocation problem is formulated under the case of coexisting multiple active eavesdroppers and multiple passive eavesdroppers. The problem can be solved similar to the single active eavesdropper case. Due to space limitation, the derivation is omitted here.

\section{Numerical Results}
In this section, numerical results are presented to illustrate the performance of the proposed secure transmission scheme for different system parameters. Without loss of generality, we assume $R_b=8$ bit/s/Hz in this section. The total transmission power  $P_{max}$  and transmission power at the active eavesdropper $P_{E_a}$ are, respectively, normalized by the noise variance, and denoted as $\bar{P}_{max}$  and ${\bar{P}}_{E_a}$ in the following simulations, i.e., $\bar{P}_{max}={P_{max}}/{N_0}$ and $\bar{P}_{E{a}}={P_{E{a}}}/{N_0}$. In addition, $\bar\gamma_{A,B}=\sigma^2_{A,B}/{N_0}$, $\bar\gamma_{A,E_a}=\sigma^2_{A,E_a}/{N_0}$, $\bar\gamma_{A,E_k}=\sigma^2_{A,E_k}/{N_0}$, and $ \bar\gamma_{E_a,B}=\sigma^2_{E_a,B}/{N_0}$ denote the normalized channel quality of Alice $\rightarrow$ Bob, Alice $\rightarrow$ $E_a$, Alice $\rightarrow$ $E_k$, and $E_a$ $\rightarrow$ Bob links, respectively. Similarly, $\bar\gamma_{A,B}=\sigma^2_{A,B}/{N_0}$, $\bar\gamma_{A,E_a}=\sigma^2_{A,E_a}/{N_0}$, $\bar\gamma_{A,E_k}=\sigma^2_{A,E_k}/{N_0}$ denote the normalized channel quality of the Jammer $\rightarrow$ Bob, Jammer $\rightarrow$ $E_a$, and Jammer $\rightarrow$ $E_k$ links, respectively.
\begin{figure}
\centering
\includegraphics[width=0.45\textwidth]{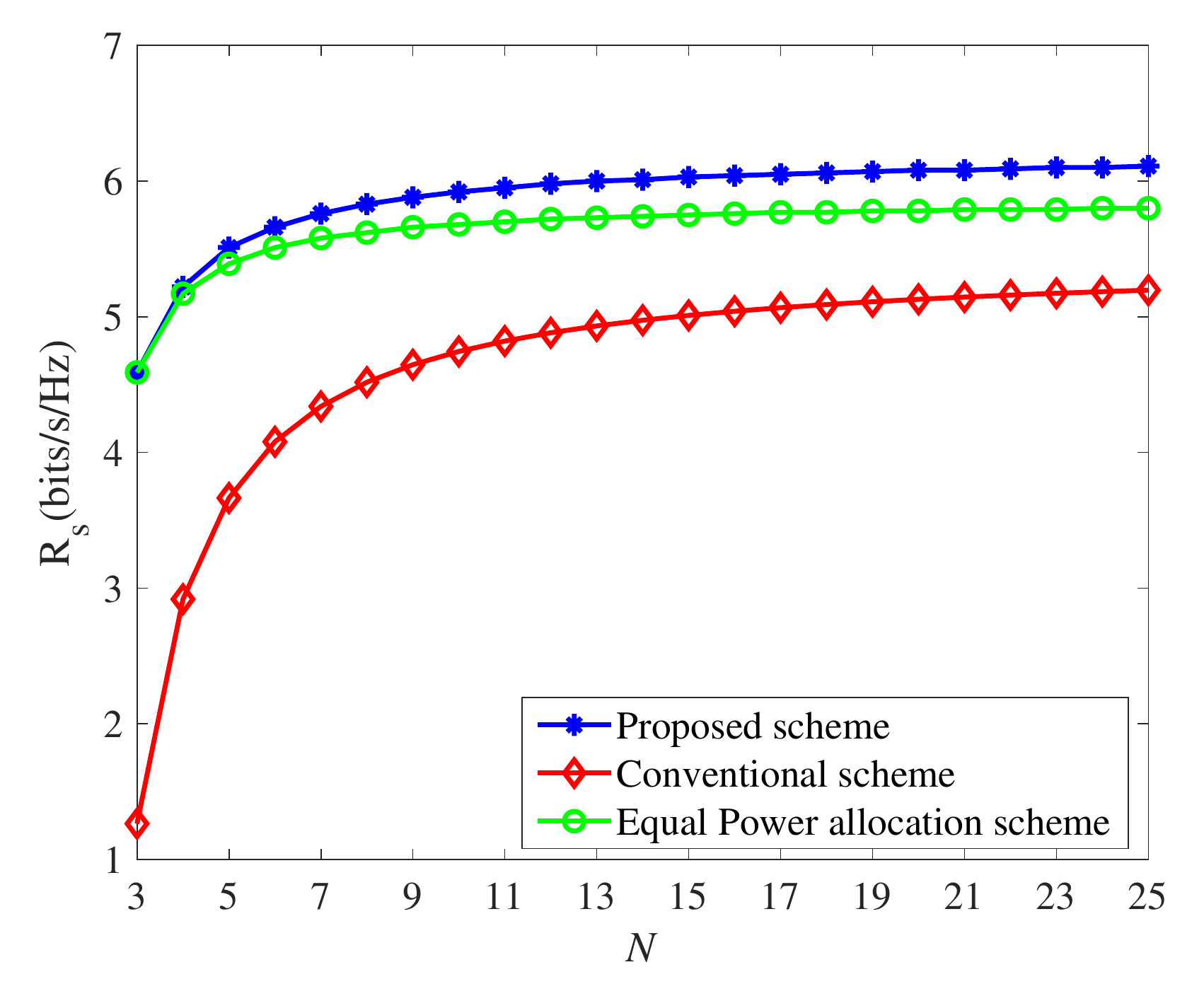}
\caption{The maximum secrecy rate against $N$. $K=1$, ${\bar\gamma}_{A,E_a}=3dB$, ${\bar\gamma}_{J,E_a}=7dB$, ${\bar\gamma}_{A,B}=10dB$, $\delta=0.1$, ${\bar{P}}_{\max}=40dB$, ${\bar{P}}_{E_a}=10dB$, ${\bar\gamma}_{E_a,B}=3dB$, and $\varepsilon=10^{-2}$.} \label{Fig2}
\end{figure}

Figure 2 compares the maximum secrecy rate for the proposed scheme and the traditional AN scheme, where the power $P_{\max}-P_A$ is uniformly allocated to the $N-1$ dimensions \cite{XZhou,NYang3,NYang4}. To illustrate the effect of the active eavesdropper, we assume one active eavesdropper and $K$ passive eavesdroppers in the proposed schemes and only $K$+1 passive eavesdroppers in the conventional scheme. In addition, we aasume an equal power allocation scheme with the two-fold zero forcing beamforming as another benchmark. To compare the three schemes explicitly and fairly, we assume that  ${\bar\gamma}_{A,E_a}={\bar\gamma}_{A,E_k}$ and ${\bar\gamma}_{J,E_a}={\bar\gamma}_{J,E_k}$  when the active eavesdropper and the passive eavesdroppers coexist. In addition, we assume that the SOP in the conventional scheme is less than $\zeta=\varepsilon^2$. It can be seen that the achievable maximum secrecy rate for the proposed scheme is larger than its values for both conventional scheme and equal power allocation scheme. This shows the superiority of our proposed algorithm. Moreover, the three curves become almost flat for larger $N$. This shows that for a given beamforming and power allocation scheme, the impact of $N$ on secrecy rate is limited when $N$ is large.

\begin{figure}
\centering
\includegraphics[width=0.45\textwidth]{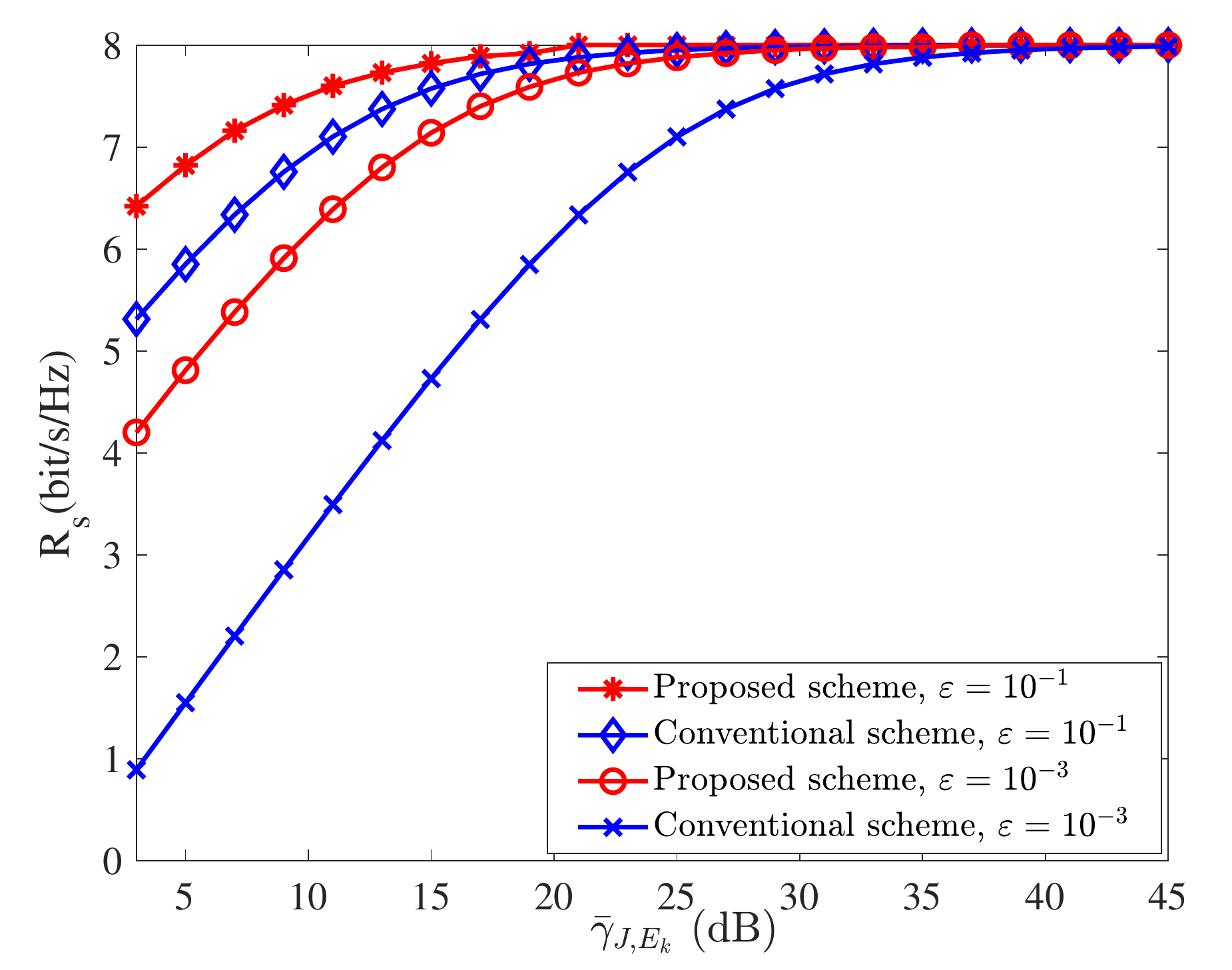}
\caption{The maximum secrecy rate against $\bar{\gamma}_{J,E_k}$.  $N=4$, $K=3$, $\bar{\gamma}_{A,E_a}=3dB$, $\bar{\gamma}_{A,B}=10dB$, $\delta=0.1$, $\bar{P}_{E_a}=10dB$, $\bar{\gamma}_{E_a,B}=3dB$, and $\bar{P}_{\max}=40dB$.} \label{Fig3}
\end{figure}

\begin{figure}
\centering
\includegraphics[width=0.45\textwidth]{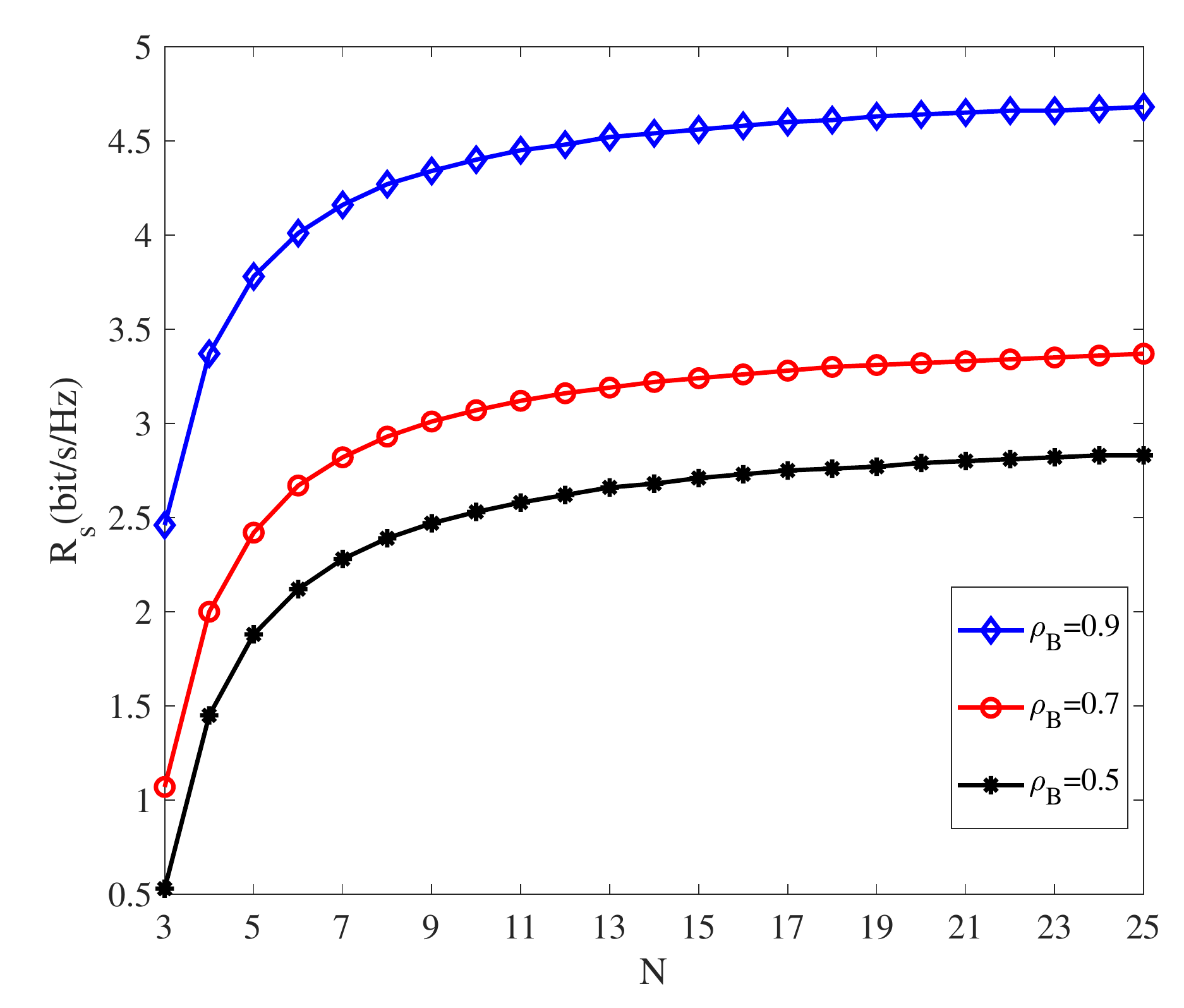}
\caption{The impact of  $\rho_{B}$ on secrecy rate. $K=5$, ${\bar\gamma}_{A,E_a}=2dB$, ${\bar\gamma}_{A,E_k}=2dB$, ${\bar\gamma}_{J,B}=2dB$, $\varepsilon=10^{-2}$, ${\bar\gamma}_{J,E_k}=10dB$, ${\bar\gamma}_{A,B}=20dB$, $\delta=0.1$, ${\bar{P}}_{E_a}=10dB$, ${\bar\gamma}_{E_a,B}=3dB$, and ${\bar{P}}_{\max}=40dB$.} \label{Fig4}
\end{figure}
The secrecy rate for different values of ${\bar\gamma}_{J,E_k}$ is illustrated in Fig. 3. It is clear that with different SOP constraints, the secrecy rate for the proposed scheme is still better than its value for the conventional scheme. Moreover, the gap between the two schemes becomes large when the SOP constraint decreases. The advantage of the proposed scheme is obvious when the SOP constraint is more strict. In addition, we can see that the secrecy rate for the four cases approaches $R_b$  when ${\bar\gamma}_{J,E_k}$ goes to infinity. This is because when ${\bar\gamma}_{J,E_k}={\bar\gamma}_{J,E_a}$ and both of them parameters approach infinity, all eavesdroppers in the four curves cannot correctly decode the information, and the secrecy performance can be guaranteed regardless of the AN power allocation between the active eavesdropper and the passive eavesdroppers. This corresponds to the extreme scenario when all the eavesdroppers are close to the jammer and far away from Alice.
\begin{figure}
\centering
\includegraphics[width=0.45\textwidth]{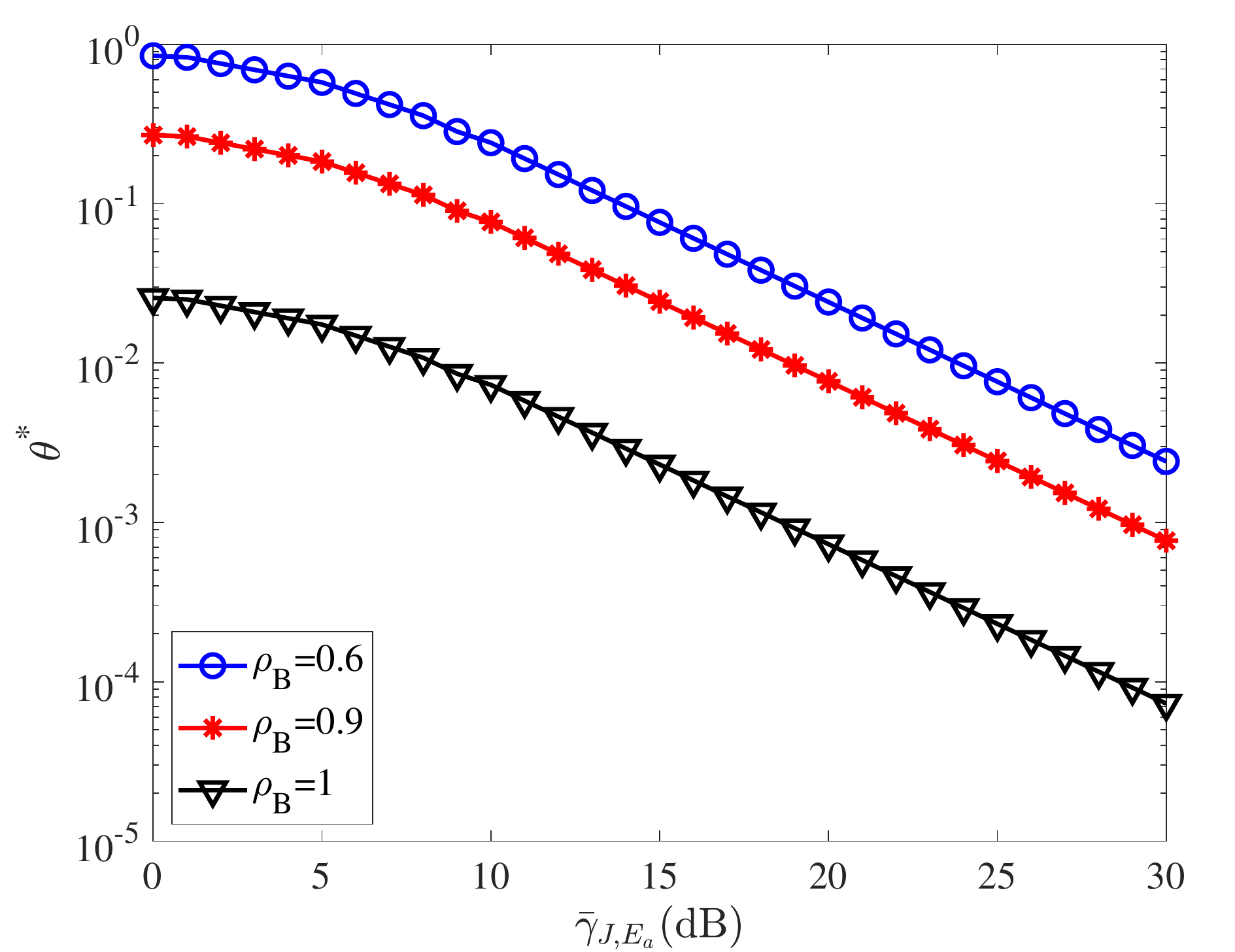}
\caption{The optimal power ratio $\theta^*$ for maximum secrecy rate against ${\bar\gamma}_{J,E_a}$.  $N=6$, $K=5$, ${\bar\gamma}_{A,E_a}=10dB$, ${\bar\gamma}_{A,E_k}=2dB$, ${\bar\gamma}_{J,B}=2dB$, ${\bar\gamma}_{J,E_k}=5dB$, ${\bar\gamma}_{A,B}=20dB$, $\varepsilon=10^{-2}$, $\delta=0.2$, $\bar{P}_{E_a}=10dB$, ${\bar\gamma}_{E_a,B}=3dB$,  and $\bar{P}_{\max}=30dB$.} \label{Fig5}
\end{figure}

The impact of $\rho_{B}$ on the secrecy rate is presented in Fig. 4, where the maximum achieved secrecy rate increases with $\rho_{B}$. A small $\rho_{B}$ means that more AN will be leaked to Bob, and more power is required by Alice to satisfy the transmission quality requirement. As a result, the power allocated to AN is decreased, the interference at all eavesdroppers is decreased, and the maximum supported secrecy rate is decreased. In addition, similar to the perfect CSI case, the secrecy rate increases with $N$.

The impact of $\rho_{B}$ on the optimal power ratio $\theta^*$ is plotted in Fig. 5, where we observe that $\theta^*$ decreases with ${\bar\gamma}_{J,{E_a}}$, because the SOP for the active eavesdropper decreases with ${\bar\gamma}_{J,{E_a}}$. The Optimal $\theta^*$  is determined by SOPs for both the active and passive eavesdroppers. In this figure, the SOP for the passive eavesdroppers is always satisfied. When ${\bar\gamma}_{J,{E_a}}$ is large, little power is required by the active eavesdropper to satisfy the SOP constraint and $\theta^*$ decreases. In addition, $\theta^*$ decreases with $\rho_{B}$ because a large $\rho_{B}$ means that little power is leaked to Bob and the power allocated to Alice is decreased. As a result, more power can be allocated to AN and $\theta^*$ decreases. Moreover, when $\rho_B=1$, $\theta^*$ always decreases with ${\bar\gamma}_{J,E_a}$ and approaches zero, but can not equal zero, which coincides with \emph{Remark 2.}

\begin{figure}
\centering
\includegraphics[width=0.5\textwidth]{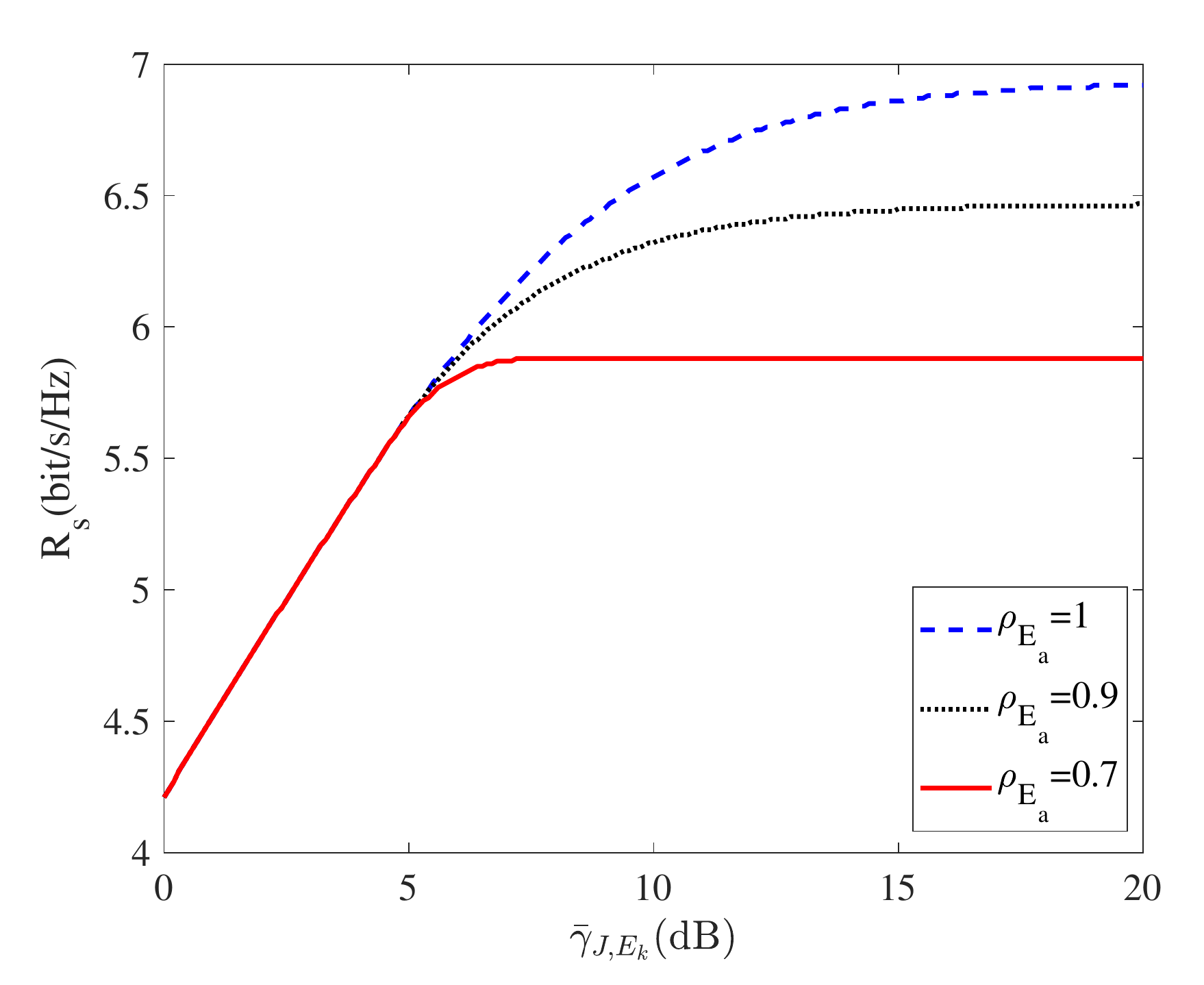}
\caption{The secrecy rate against ${\bar\gamma}_{J,E_k}$.  $N=5$, $K=3$, ${\bar\gamma}_{A,E_a}=5dB$, ${\bar\gamma}_{A,E_k}=5dB$, ${\bar\gamma}_{J,B}=2dB$, ${\bar\gamma}_{J,E_a}=3dB$, ${\bar\gamma}_{A,B}=15dB$, $\delta=0.1$, $\varepsilon=10^{-2}$, $P_{E_a}=10dB$, ${\bar\gamma}_{E_a,B}=3dB$,  and  ${\bar{P}}_{\max}=35dB$.} \label{Fig6}
\end{figure}

Figure 6 illustrates the impact of $\rho_{E_a}$ on the secrecy rate for different values of ${\bar\gamma}_{J,E_k}$. When ${\bar\gamma}_{J,E_k}$ is small, the three curves converge for different values of $\rho_{E_a}$, but when ${\bar\gamma}_{J,E_k}$ is greater than 6 dB, these three curves begin to separate. When ${\bar\gamma}_{J,E_k}$ is greater than 15 dB, the three curves become flat because the secrecy rate is determined by the SOPs for both the active and passive eavesdroppers. When the active eavesdropper is far away from the jammer, the secrecy rate is only determined by the SOP for the passive eavesdroppers. Hence, the secrecy rates for the three different correlation coefficients are the same due to poor channel quality between the jammer and the passive eavesdropper. On the contrary, the secrecy rate is determined by the SOP for the active eavesdropper when the channel quality between the jammer and the passive eavesdroppers is good. In this case, though the AN towards the passive eavesdroppers leaks to the active eavesdropper, the MRT for the active eavesdropper cannot be guaranteed due to imperfect ${{\tilde{ \bf{g}}}_{{E_a}}}$. Hence, the secrecy rate increases with $\rho_{E_a}$ at large ${\bar\gamma}_{J,E_k}$. These results support \emph{Lemma 2}.

\begin{figure}
\centering
\includegraphics[width=0.50\textwidth]{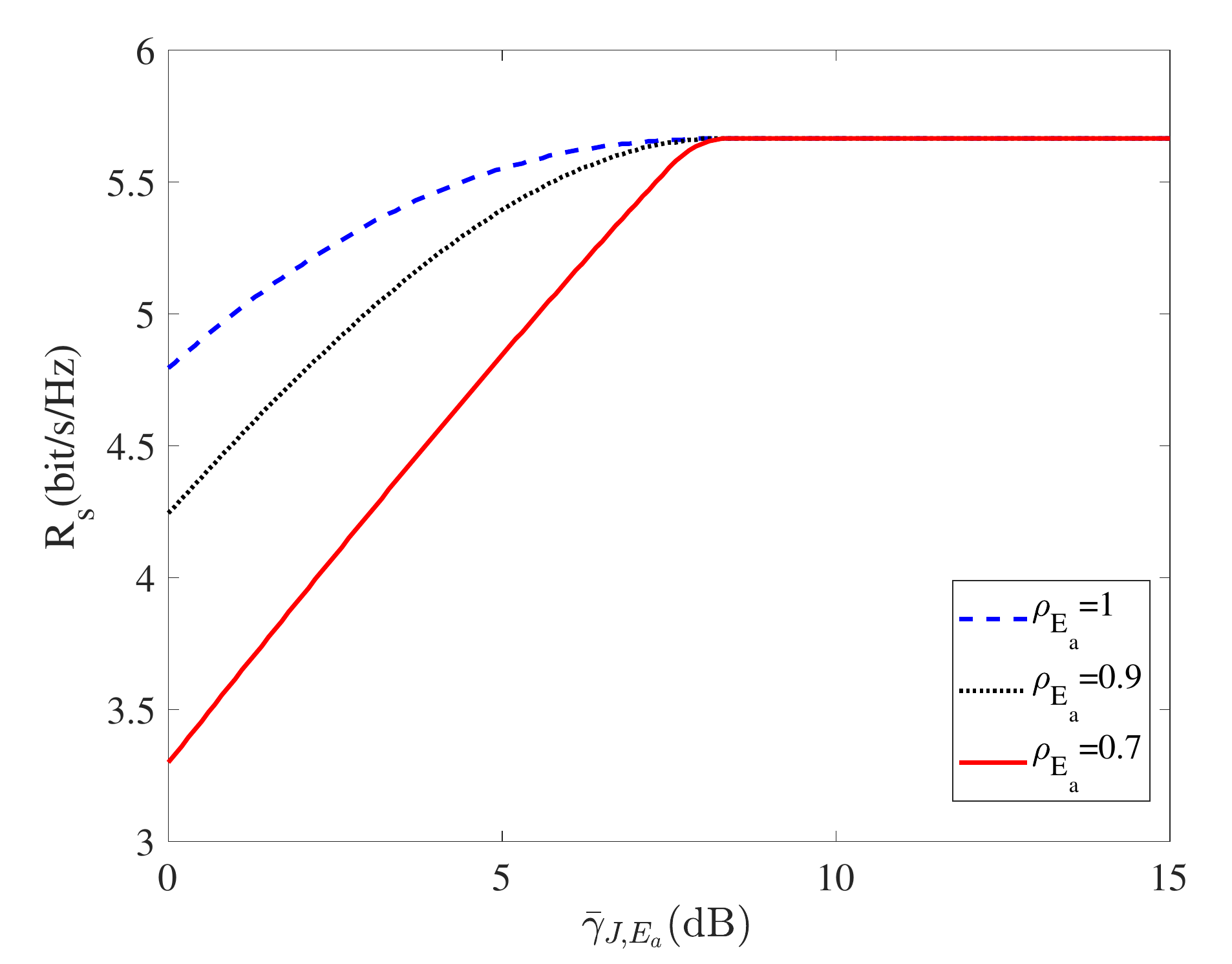}
\caption{The secrecy rate against ${\bar\gamma}_{J,E_a}$.  $N=5$, $K=3$, ${\bar\gamma}_{A,E_a}=10dB$, ${\bar\gamma}_{A,E_k}=5dB$, ${\bar\gamma}_{J,B}=2dB$, ${\bar\gamma}_{J,E_k}=5dB$, ${\bar\gamma}_{A,B}=15dB$, $\varepsilon=10^{-2}$, $\delta=0.1$, ${\bar{P}}_{E_a}=10dB$, ${\bar\gamma}_{E_a,B}=3dB$,  and  ${\bar{P}}_{\max}=35dB$.} \label{Fig7}
\end{figure}

The secrecy rate for different values of ${\bar\gamma}_{J,E_a}$ is plotted in Fig. \ref{Fig7}. Different from  Fig. \ref{Fig6}, when $\sigma^2_{J,E_a}$ is small, the secrecy rate increases with $\rho_{E_a}$, but the three curves converge when ${\bar\gamma}_{J,E_a}$ is larger than 8 dB. The reason for this behavior is similar to Fig. \ref{Fig6}. Since for large ${\bar\gamma}_{J,E_a}$, $\rho_{E_a}$ has no relationship with the SOP for the passive eavesdroppers and the secrecy rate is only determined by the SOP for the passive eavesdropper, the three curves have the same maximum secrecy rate. By contrast, when ${\bar\gamma}_{J,E_a}$ is small, the SOP for the active eavesdropper determines the maximum secrecy rate.

\begin{figure}
\centering
\includegraphics[width=0.45\textwidth]{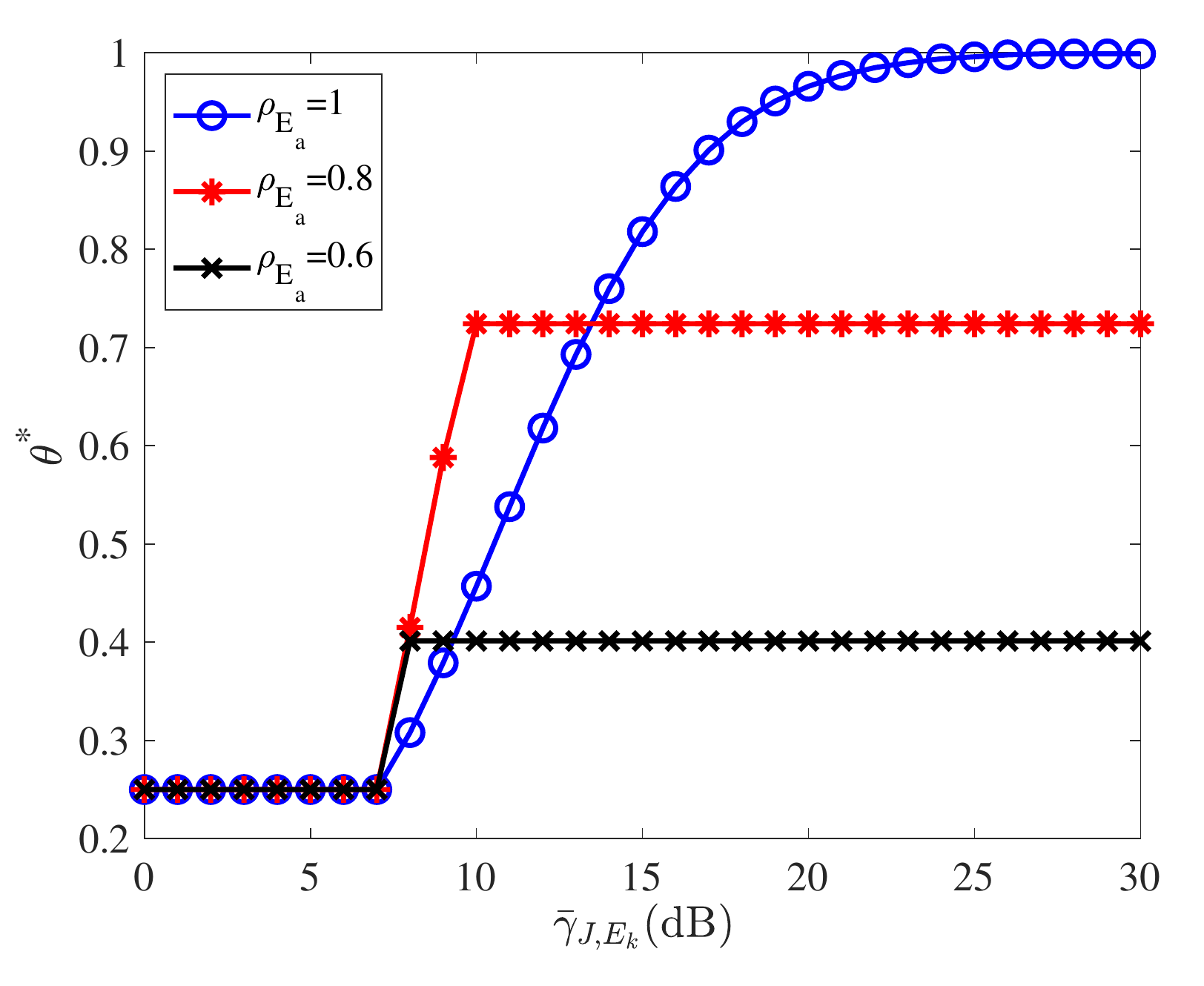}
\caption{ Optimal ${\theta^*}$ against ${\bar\gamma}_{J,E_k}$. $N=5$, $K=4$, ${\bar\gamma}_{A,E_a}=3dB$, ${\bar\gamma}_{A,E_k}=5dB$, ${\bar\gamma}_{J,B}=2dB$, ${\bar\gamma}_{J,E_a}=3dB$, ${\bar\gamma}_{A,B}=15dB$, $\varepsilon=10^{-2}$, $\delta=0.1$, ${\bar{P}}_{E_a}=10dB$, ${\bar\gamma}_{E_a,B}=3dB$, and ${\bar{P}}_{\max}=35dB$.} \label{Fig8}
\end{figure}

The optimal ${\theta^*}$ is plotted against ${\bar\gamma}_{J,E_k}$ in Fig. \ref{Fig8}. When ${\bar\gamma}_{J,E_k}$ is small, the optimal $\theta^*$  for different $\rho_{E_a}$ have the same value of $1/(N-1)=0.25$. As in this case, the SOP for the active eavesdropper is always less than that for the passive eavesdroppers. As a result, the maximum secrecy rate is determined by the SOP for the passive eavesdroppers, and the optimal $\theta^*$ for the maximum secrecy rate is $1/(N-1)$. In addition, ${\theta^*}$ increases with  ${\bar\gamma}_{J,E_k}$ because in this ${\bar\gamma}_{J,E_k}$ range, the optimal $\theta^*$ satisfies the SOP constraints for both the active and passive eavesdroppers. On the one hand, when ${\bar\gamma}_{J,E_k}$ goes to infinity and $\rho_{E_a}\neq 1$, we can see that the optimal $\theta^*$ becomes flat. In this case, the maximum secrecy rate is determined by the SOP for the active eavesdropper, and the optimal $\theta^*$ is the optimal solution of SOP for the active eavesdropper. Note that there are sudden changes in the curves for $\rho_{E_a}=0.6$ and $\rho_{E_a}=0.8$, which implies  that with the increase of ${\bar\gamma}_{J,E_k}$, the maximum achieved secrecy rate is dominated by the SOP of the active eavesdropper. On the other hand, when ${\bar\gamma}_{J,E_k}$ goes to infinity and $\rho_{E_a}=1$, the optimal $\theta^*$ approaches unity but cannot equal unity. In this case, the maximum secrecy rate is determined by the SOP for the active eavesdropper, which is a monotonous decreasing function of $\theta$. As a result, $\theta^*$ approaches unity. These results agree with \emph{Remark 2} and \emph{Remark 4 }in Section IV.

\section{Conclusion}
We proposed a two-fold zero-forcing jamming and beamforming scheme for secrecy transmission in the presence of both active and passive eavesdroppers. By taking the instantaneous CSI between the jammer and the active eavesdropper into account, the proposed scheme can achieve better secrecy performance than the conventional AN scheme with beamforming and can be adopted in practice with low complexity. In addition, imperfect CSI between the jammer and the legitimate receiver will do more harm to the achievable secrecy rate than imperfect CSI between the jammer and the active eavesdropper. Moreover, we generalized the proposed scheme to the multiple active eavesdroppers case when there is enough number of antennas at the jammer.

\begin{appendices}
\section{SOP for Active Eavesdropper With Imperfect CSI ${{\tilde{\bf{g}}}_{E_a}}$}
Due to imperfect CSI ${{\tilde{\bf{g}}}_{E_a}}$, the SNR at the active eavesdropper is written as
\begin{align}\label{imperfectEA}\
{\tilde \gamma _{{E_a}}} &= \frac{{{P_A}{{\left| {{h_{A,{E_a}}}} \right|}^2}}}{{{P_{{J_A}}}{{\left| {{{{{\tilde{\mbox{\small \bf {W}}}}}}^H}_{{E_a}}{{\bf{g}}_{_{{E_a}}}}} \right|}^2} + \frac{{{P_{{J_p}}}}}{{N - 2}}\left\| {{{{\bf{\tilde W}}}^H}_{{E_p}}{{\bf{g}}_{_{{E_a}}}}} \right\| }}
\end{align}
where ${{{{{{{\tilde{\mbox{\small \bf {W}}}}}}^H}}_{{E_a}}}}= \frac{{{{{\bf{\tilde g}}}^H}_{{E_a}}{{\left( {{{\bf{I}}_N} - \frac{{{{\bf{g}}_B}{{\bf{g}}^H}_B}}{{{{\left\| {{{\bf{g}}_B}} \right\|}^2}}}} \right)}^H}}}{{\left\| {\left( {{{\bf{I}}_N} - \frac{{{{\bf{g}}_B}{{\bf{g}}^H}_B}}{{{{\left\| {{{\bf{g}}_B}} \right\|}^2}}}} \right){{{\bf{\tilde g}}}^*}_{{E_a}}} \right\|}}$.  Since ${\left( {{{\bf{I}}_N} - \frac{{{{\bf{g}}_B}{{\bf{g}}^H}_B}}{{{{\left\| {{{\bf{g}}_B}} \right\|}^2}}}} \right)}$ is an idempotent matrix,\\
 ${{{\left( {{{\bf{I}}_N} - \frac{{{{\bf{g}}_B}{{\bf{g}}^H}_B}}{{{{\left\| {{{\bf{g}}_B}} \right\|}^2}}}} \right)}^H}=\left( {{{\bf{I}}_N} - \frac{{{{\bf{g}}_B}{{\bf{g}}^H}_B}}{{{{\left\| {{{\bf{g}}_B}} \right\|}^2}}}} \right)}$ and $\left( {{{\bf{I}}_N} - \frac{{{{\bf{g}}_B}{{\bf{g}}^H}_B}}{{{{\left\| {{{\bf{g}}_B}} \right\|}^2}}}} \right) = {\left( {{{\bf{I}}_N} - \frac{{{{\bf{g}}_B}{{\bf{g}}^H}_B}}{{{{\left\| {{{\bf{g}}_B}} \right\|}^2}}}} \right)^2}$. Thus, ${{{{{{{\tilde{\mbox{\small \bf {W}}}}}}^H}}_{{E_a}}}}{{{\bf{\tilde g}}}_{{E_a}}}$ can be rewritten as \cite{YMa,ZHu, YYang}
\begin{align}\label{impedence}\
{{{{{{{\tilde{\mbox{\small\bf{W}}}}}}^H}}_{{E_a}}}}{{{\bf{\tilde g}}}_{{E_a}}} &= \frac{{{{{\bf{\tilde g}}}^H}_{{E_a}}{{\left( {{{\bf{I}}_N} - \frac{{{{\bf{g}}_B}{{\bf{g}}^H}_B}}{{{{\left\| {{{\bf{g}}_B}} \right\|}^2}}}} \right)}^H}\left( {{{\bf{I}}_N} - \frac{{{{\bf{g}}_B}{{\bf{g}}^H}_B}}{{{{\left\| {{{\bf{g}}_B}} \right\|}^2}}}} \right)}}{{\left\| {\left( {{{\bf{I}}_N} - \frac{{{{\bf{g}}_B}{{\bf{g}}^H}_B}}{{{{\left\| {{{\bf{g}}_B}} \right\|}^2}}}} \right){{{\bf{\tilde g}}}_{{E_a}}}} \right\|}}{{{\bf{\tilde g}}}_{{E_a}}}\nonumber  \\
 &= \left\| {\left( {{{\bf{I}}_N} - \frac{{{{\bf{g}}_B}{{\bf{g}}^H}_B}}{{{{\left\| {{{\bf{g}}_B}} \right\|}^2}}}} \right){{{\bf{\tilde g}}}_{{E_a}}}} \right\|.
\end{align}
Substituting \eqref{impedence} into \eqref{imperfectEA}, we can rewrite the SNR at the active eavesdropper as
\begin{align}\label{imperfectEA2}\
{\tilde \gamma _{{E_a}}}= \frac{{{P_A}{{\left| {{h_{A,{E_a}}}} \right|}^2}}}{{{P_{{J_A}}}{{\left| {{\rho _{{E_a}}}z + {{{{\tilde {\mbox {\small\bf{W}}}}}}^H}_{{E_a}}{{\bf{e}}_{{E_a}}}} \right|}^2} + \frac{{{P_{{J_p}}}}}{{N - 2}}{{\left\| {{{{\bf{\tilde W}}}^H}_{{E_p}}{{\bf{e}}_{{E_a}}}} \right\|}^2}}}
\end{align}
where $z={\left\| {\left( {{{\bf{I}}_N} - \frac{{{{\bf{g}}_B}{{\bf{g}}^H}_B}}{{{{\left\| {{{\bf{g}}_B}} \right\|}^2}}}} \right){{{\bf{\tilde g}}}_{{E_a}}}} \right\|}$. In \eqref{imperfectEA2}, conditioned on $z$, then ${\rho _{{E_a}}}z + {{{{\tilde {\mbox {\small\bf{W}}}}}}^H}_{{E_a}}{{\bf{e}}_{{E_a}}}$ follows $\mathcal{CN}\left({\rho _{{E_a}}}z,
\frac{{\left( {1 - {\rho ^2}_{{E_a}}} \right){\sigma ^2}_{J,{E_a}}}}{2} \right)$. When $ {\rho _{{E_a}}}= 0$, for $\|{{{{\tilde {\mbox {\small\bf{W}}}}}}^H}_{{E_a}}\|=1$, ${\left| {{\rho _{{E_a}}}z + {{{{\tilde {\mbox {\small\bf{W}}}}}}^H}_{{E_a}}{{\bf{e}}_{{E_a}}}} \right|^2}$ follows an exponential distribution, $\exp\left({\sigma ^2_{J,E_a}}\right)$. The MGF of ${\left| {{\rho _{{E_a}}}z + {{{{\tilde {\mbox {\small\bf{W}}}}}}^H}_{{E_a}}{{\bf{e}}_{{E_a}}}} \right|^2}$ is given by
 \begin{align}\label{mgfEA1}\
 \phi_{\lambda_5} \left( s \right) ={\frac{1}{{1 + s{{\sigma ^2}_{J,{E_a}}}}}}.
\end{align}
By contrast, when $ {\rho _{{E_a}}}\neq 0$, $\lambda_5={\left| {{\rho _{{E_a}}}z + {{{{\tilde {\mbox {\small\bf{W}}}}}}^H}_{{E_a}}{{\bf{e}}_{{E_a}}}} \right|^2}$ follows a non-central chi-square distribution having two degrees of freedom. The conditional MGF of ${\left| {{\rho _{{E_a}}}z + {{{{\tilde {\mbox {\small\bf{W}}}}}}^H}_{{E_a}}{{\bf{e}}_{{E_a}}}} \right|^2}$ is given by \cite{Simon,IS}
 \begin{align}\label{mgfEA2}\
\phi_{\lambda_5|z} \left( {s\left| z \right.} \right) &= {\frac{1}{{1 + s\left( {\left( {1 - {\rho ^2}_{{E_a}}} \right){\sigma ^2}_{J,{E_a}}} \right)}}} \nonumber \\
&\times \exp \left( {\frac{{ - s{\rho ^2}_{{E_a}}{z^2}}}{{1 + s\left( {\left( {1 - {\rho ^2}_{{E_a}}} \right){\sigma ^2}_{J,{E_a}}} \right)}}} \right).
\end{align}
Note that $z^2$ follows a central chi-square distribution having $2(N-1)$ degrees of freedom, and the MGF of $z^2$  is $\mathbb{E}\left( { - {z^2}s} \right) = {\left( {\frac{1}{{1 + {\sigma ^2}_{J,{E_a}}s}}} \right)^{N - 1}}$. Integrating \eqref{mgfEA2} w.r.t. $z$, we obtain the MGF of $\lambda_5$ as
 \begin{align}\label{mgfEA3}\
\phi_{\lambda_5} \left( s \right) &=\mathbb{E}\left(\phi_{\lambda_5|z} \left( {s\left| z \right.} \right)\right)\nonumber\\
&=\frac{{{{\left( {1 + \left( {1 - \rho _{E_a}^2} \right){\sigma ^2}_{J,{E_a}}s} \right)}^{N - 2}}}}{{{{\left( {1 + {\sigma ^2}_{J,{E_a}}s} \right)}^{N - 1}}}}.
\end{align}
Since $\lambda_6=\frac{{2{{\left\| {{{{\bf{\tilde W}}}^H}_{{E_p}}{{\bf{e}}_{{E_a}}}} \right\|}^2}}}{{\left( {1 - {\rho ^2}_{{E_a}}} \right){\sigma ^2}_{J,{E_a}}}}$ follows a central chi-square distribution  ${\chi ^2}\left( {2\left( {{N} - 2} \right)} \right)$, the MGF of $\lambda_6$ is given by
 \begin{align}\label{mgfEA5}\
\phi_{\lambda_6} \left( s \right) ={\left( {\frac{{1}}{{1 +s}}} \right)^{N - 2}}.
 \end{align}
According to \eqref{imperfectEA2}, \eqref{mgfEA1}, \eqref{mgfEA3}, and \eqref{mgfEA5}, the CDF of ${\tilde \gamma _{{E_a}}}$ is calculated as
 \begin{align}\label{CDFEA2}\
&F_{\tilde \gamma _{{E_a}}}\left( y \right)=\Pr \left( {{\gamma _{{E_a}}} \le y} \right) \nonumber \\
 & =1- \mathbb{E}\left( {\exp \left( { - \frac{{{P_{{J_A}}}{\lambda _5} + \frac{{{P_{{J_P}}}\left( {1 - {\rho ^2}_{{E_a}}} \right){\sigma ^2}_{J,{E_a}}{\lambda _6}}}{{2\left( {N - 2} \right)}}}}{{{P_A}{\sigma ^2}_{A,{E_a}}}}} \right)} \right)\nonumber \\
 &= 1-\mathbb{E}\left( {\exp \left( { - \frac{{{P_{{J_A}}}y{\lambda _5}}}{{{P_A}{\sigma ^2}_{A,{E_a}}}}} \right)} \right)\nonumber \\
 &\quad\times\mathbb{E}\left( \exp \left(-{\frac{{{P_{{J_P}}}\left( {1 - {\rho ^2}_{{E_a}}} \right){\sigma ^2}_{J,{E_a}}y{\lambda _6}}}{{2{P_A}{\sigma ^2}_{A,{E_a}}\left( {N - 2} \right)}}}\right) \right)\nonumber  \\
&=1 - {\phi _{{\lambda _5}}}\left( {  \frac{{{P_J}_{_A}y}}{{{P_A}{\sigma ^2}_{A,{E_a}}}}} \right){\phi _{{\lambda _6}}}\left( {  \frac{{{P_J}_{_p}\left( {1 - {\rho ^2}_{{E_a}}} \right){\sigma ^2}_{J,{E_a}}y}}{{2{P_A}{\sigma ^2}_{A,{E_a}}\left( {N - 2} \right)}}} \right)\nonumber  \\
 &= 1-\frac{{{{\left( {1 + \frac{{{P_{{J_A}}}\left( {1 - {\rho ^2}_{{E_a}}} \right){\sigma ^2}_{J,{E_a}}y}}{{{P_A}{\sigma ^2}_{A,{E_a}}}}} \right)}^{N - 2}}}}{{{{\left( {1 + \frac{{{P_{{J_A}}}{\sigma ^2}_{J,{E_a}}y}}{{{P_A}{\sigma ^2}_{A,{E_a}}}}} \right)}^{N - 1}}}}\nonumber \\
&\quad \times{\left( {1 + \frac{{{P_{{J_P}}}\left( {1 - {\rho ^2}_{{E_a}}} \right)y{\sigma ^2}_{J,{E_a}}}}{{{P_A}{\sigma ^2}_{A,{E_a}}\left( {N - 2} \right)}}} \right)^{\left( {2 - N} \right)}}.
 \end{align}
In addition, the SOP for the active eavesdropper is given by
 \begin{align}\label{pso1imperfect2}\
 {p_{{{so}_1}}} = 1 - {F_{{{\tilde \gamma }_{{E_a}}}}}\left( {{2^{{R_b} - {R_s}}} - 1} \right).
 \end{align}
 Substituting \eqref{CDFEA2} into \eqref{pso1imperfect2}, we can obtain the exact expression of ${p_{{{so}_1}}}$ in \eqref{pso1imperfect}.

\section{Impact of $\rho_{E_a}$, $R_s$, $P_A$, and $\theta$ on $p_{so_1}$ With Imperfect CSI ${{\tilde{\bf{g}}}_{E_a}}$}
According to \eqref{pso1imperfect}, the derivative of $p_{so_1}$ w.r.t. $\rho_{E_a}$ is given by
 \begin{align}\label{pso1rhoea}\
\frac{{\partial {p_{s{o_1}}}}}{{\partial {\rho _{{E_a}}}}} &=  - \frac{2{\rho _{{E_a}}}\theta \alpha {\left( {N - 1 + \left( {1 - {\rho ^2}_{{E_a}}} \right)\alpha } \right)}}{{{{\left( {1 + \theta \alpha } \right)}^{N - 1}}}}\nonumber\\
&\quad \times\frac{{{\left( {N - 2} \right)}^{N - 2}}{{\left( {1 + \theta \left( {1 - {\rho ^2}_{{E_a}}} \right)\alpha } \right)}^{N - 3}}}{{{\left( {N - 2 + \left( {1 - \theta } \right)\left( {1 - {\rho ^2}_{{E_a}}} \right)\alpha } \right)}^{N - 1}}}.
 \end{align}
We find from \eqref{pso1rhoea} that $\frac{{\partial {P_{s{o_1}}}}}{{\partial {\rho _{{E_a}}}}}<0$, which  means that ${P_{s{o_1}}}$ decreases with ${\rho _{{E_a}}}$. Similarly, we can verify that the derivative of $p_{so_1}$ w.r.t. $P_A$, $\frac{{\partial {p_{so_1}}}}{{\partial {P_A}}}>0$, and the derivative of $p_{so_1}$ w.r.t. $R_s$, $\frac{{\partial {p_{so_1}}}}{{\partial {R_s}}}> 0$. In addition, the derivative of $p_{so_1}$ w.r.t. $\theta$ can be written as
 \begin{align}\label{pso1thetaimperfect}\
\frac{{\partial {p_{so_1}}}}{{\partial \theta }} = A\left( \theta  \right)J\left( \theta  \right)
 \end{align}
 where
 \begin{align}\label{pso1thetaimperfect}\
A\left( \theta  \right) &= \frac{{\left( {1 - \rho _{{E_a}}^2} \right){\alpha ^2}{{\left( {1 + \theta \left( {1 - \rho _{{E_a}}^2} \right)\alpha } \right)}^{N - 3}}}}{{{{\left( {1 + \theta \alpha } \right)}^N}}} \nonumber \\
& \times {\left( {1 + \frac{{\left( {1 - \theta } \right)\left( {1 - \rho _{{E_a}}^2} \right)\alpha }}{{N - 2}}} \right)^{1 - N}}
 \end{align}
and
\begin{align}\label{Jtheta}\
J\left( \theta  \right) &= {\theta ^2}\left( {1 - \rho _{{E_a}}^2} \right)\alpha \frac{{N - 1}}{{N - 2}} + \theta \left( {\frac{{N - 1 - \left( {1 - \rho _{{E_a}}^2} \right)\alpha }}{{N - 2}}} \right) \nonumber \\
&- \frac{{\left( {N - 1} \right)\rho _{{E_a}}^2}}{{\left( {1 - \rho _{{E_a}}^2} \right)\alpha }} - \frac{{N - 1}}{{N - 2}} + \left( {1 - \rho _{{E_a}}^2} \right).
 \end{align}
Hence $A\left( \theta  \right)>0$ and $J\left( \theta  \right)$ determines whether $\frac{{\partial {P_{so1}}}}{{\partial \theta }}$ is greater than zero or not. Fortunately, $J\left( \theta  \right)$ is a quadratic polynomial in $\theta$. Therefore, the minimum value of $J\left( \theta  \right)$  is given by
 \begin{align}\label{minJtheta}\
\min \left( {J\left( \theta  \right)} \right)&=-\frac{{ \left( {\frac{{\left( {N - 1} \right){\rho ^2}_{{E_a}}}}{{1 - {\rho ^2}_{{E_a}}}} + \alpha \left( {{\rho ^2}_{{E_a}} + \frac{1}{{N - 2}}} \right)} \right)}}{\alpha }\nonumber \\
 &- \frac{{\left( {N - 1} \right){{\left( {1 - \frac{{\left( {1 - {\rho ^2}_{{E_a}}} \right)\alpha }}{{N - 1}}} \right)}^2}}}{{4\left( {N - 2} \right)\left( {1 - {\rho ^2}_{{E_a}}} \right)\alpha }}.
 \end{align}
It is clear that $\min \left( {J\left( \theta  \right)} \right)<0$. When $\theta  = - \frac{1}{{2\alpha \left( {1 - \rho _{{E_a}}^2} \right)}} + \frac{1}{{2\left( {N - 1} \right)}}
$, $\min \left( {J\left( \theta  \right)} \right)$ is also less than unity.  In addition, when $J\left( \theta  \right)=0$, we have
\begin{align}\label{theta1}\
\theta_1 & = \frac{1}{{2\Lambda \left(N-1 \right)}}\left( { - \left( {1 - \Lambda} \right)} \right.\nonumber \\
&\left. { - \sqrt {{{\left( {1 - \Lambda} \right)}^2} + 4\left( {\left( {N - 2} \right)\rho _{{E_a}}^2\left( {1 +\Lambda} \right) + \left( {N - 1} \right)^2} \right)} } \right)\nonumber \\
 &< \frac{1}{{2\Lambda \left(N-1 \right)}}\left( { - \left( {1 - \Lambda} \right) - \left| {1 - \Lambda} \right|} \right),
\end{align}
and
\begin{align}\label{theta2}\
\theta_2  &=\frac{1}{{2\Lambda \left(N-1 \right) }}\left( { - \left( {1 - \Lambda} \right)} \right.\nonumber \\
&\left. { + \sqrt {{{\left( {1 - \Lambda} \right)}^2} + 4\left( {\left( {N - 2} \right)\rho _{{E_a}}^2\left( {1 + \Lambda} \right) + \left( {N - 1} \right)^2 } \right)} } \right)
\end{align}
where $\Lambda=\frac{{\left( {1 - \rho _{{E_a}}^2} \right)\alpha }}{{N - 1}}$. It is clear that $\theta_1<0$ and $\theta_2>0$. Next, we will examine the range of $\theta_2$. If $\theta_2>1$, we have
\begin{align}\label{theta2large1}\
2\Lambda + 4\left( {N - 2} \right)\rho _{{E_a}}^2\left( {1 + \Lambda} \right)+ 4\left( {N - 1} \right)^2 - 3 > 0.
\end{align}
Since $\theta_1<0$ and the optimum solution of ${J\left( \theta  \right)}$ is less than unity,  $ {J\left( \theta  \right)}< 0 \left( 0\leq \theta \leq 1\right)$ when \eqref{theta2large1} is satisfied. Otherwise, $ 0<\theta_2 < 1$. In this case, $ {J\left( \theta  \right)}> 0 \left( \theta_2 \leq \theta \leq 1\right)$ and $ {J\left( \theta  \right)}< 0 \left( 0 \leq \theta \leq \theta_2 \right)$. According to the analysis above, Lemma 2 is obtained.

\end{appendices}
\bibliography{jammingpowerallocation2}
\end{document}